\documentclass[aip,reprint, nofootinbib]{revtex4-1}

\usepackage{graphicx}
\usepackage{subfigure}
\usepackage{url}

\usepackage{breakurl}
\usepackage[breaklinks]{hyperref}

\draft 

\begin{document}

\title{The Future of Quantum Computing with Superconducting Qubits}

\author{Sergey Bravyi}
\author{Oliver Dial}
\author{Jay M. Gambetta}
\author{Dar\'io Gil}
\author{Zaira Nazario}
%\altaffiliation{Author to whom correspondence should be addressed: zaira.nazario@ibm.com.}

\affiliation{IBM Quantum, IBM T.J. Watson Research Center, Yorktown Heights, NY 10598, USA}

\date{\today}

\begin{abstract}
For the first time in history, we are seeing a branching point in computing paradigms with the emergence of quantum processing units (QPUs). Extracting the full potential of computation and realizing quantum algorithms with a super-polynomial speedup will most likely require major advances in quantum error correction technology. Meanwhile, achieving a computational advantage in the near term may be possible by combining multiple QPUs through circuit knitting techniques, improving the quality of solutions through error suppression and mitigation, and focusing on heuristic versions of quantum algorithms with asymptotic speedups.
For this to happen, the performance of quantum computing hardware needs to improve and software needs to seamlessly integrate quantum and classical processors together to form a new architecture that we are calling quantum-centric supercomputing. Long term, we see hardware that exploits qubit connectivity in higher than 2D topologies to realize more efficient quantum error correcting codes, modular architectures for scaling QPUs and parallelizing workloads, and software that evolves to make the intricacies of the technology invisible to the users and realize the goal of ubiquitous, frictionless quantum computing.

\end{abstract}


\maketitle 

\section{Introduction}

The history of computing is that of advances born out of the need to perform ever more sophisticated calculations. Increasingly advanced semiconductor manufacturing processes have resulted in faster and more efficient chips—most recently the 2 nm technology node\cite{2nm}—and special accelerators like the GPU, TPU, and AI processors~\cite{reuther2020survey}
have allowed more efficient computations on larger data sets. These advances share the same model of computation dating back to 1936 with the origins of the Church-Turing thesis. Now for the first time in history, the field of computing has branched with the emergence of quantum computers, which, when scaled, promise to implement computations intractable for conventional computers—from modeling quantum mechanical systems~\cite{AbramsLloyd99} to linear algebra~\cite{HHL}, factoring~\cite{shor1994algorithms},
search~\cite{grover}, and more.  

Unlocking the full potential of quantum processors requires the implementation of computations with a large number of operations. Since quantum gates are considerably less accurate than their classical counterparts, it is strongly believed that error correction
will be necessary to realize long computations with millions or billions of gates. Accordingly, most of quantum computing platforms are designed with the long term goal of realizing 
error-corrected quantum circuits.
As the noise rate decreases below a constant architecture-dependent threshold, an arbitrarily long quantum circuit can be executed reliably by redundantly encoding each qubit and 
repeatedly measuring parity check operators to detect and correct errors. However, the number of qubits required to realize error-corrected
quantum circuits solving classically hard problems
exceeds the size of systems available today by several orders of magnitude.

Meanwhile, as the quality and number of qubits in quantum computers continue to grow, we must be able to harvest the computational power of quantum circuits available along the way. For example, a quantum processing unit (QPU) with two-qubit gate fidelity of $99.99\%$ can implement circuits with a few thousand gates to a fair degree of reliability
without resorting to error correction. 
Such circuits are strongly believed to be practically impossible to simulate classically, even with the help of modern supercomputers. This suggests the possibility that the first demonstrations of a computational {\it quantum advantage}—where a computational task of business or scientific relevance can be performed more efficiently, cost-effectively, or accurately using a quantum computer than with classical computations alone—{\it may be achieved without or with limited error correction}. 

Three central questions need to be answered for this to happen: (1) how to extract useful data from the output of noisy quantum circuits in the weak noise regime, (2) how to design quantum algorithms based on shallow circuits that can potentially solve some classically hard problems, and (3) how to improve the efficiency of quantum error-correction schemes
and use error correction more sparingly. 

These questions and our approach are discussed in detail in Section~\ref{sec:useful}.
As an illustration, 
we pick one of the simplest scientifically relevant applications of quantum computers—simulating 
time evolution of a spin chain Hamiltonian~\cite{childs2018toward}. 
We discuss state-of-the-art quantum algorithms for this problem
and highlight the cost of making time evolution circuits fault-tolerant by encoding each qubit into 
the surface code~\cite{kitaev2003fault,bravyi1998quantum},
considered the best fit for hardware with two-dimensional qubit connectivity.
For problem sizes of practical interest, error correction increases
the size of quantum circuits by nearly six orders of magnitude, making it prohibitively expensive for near-term
QPUs (see Section~\ref{sec:EC}). 

Question (1) is approached in Sections~\ref{sec:EM} and~\ref{sec:CK} through quantum error mitigation~\cite{temme2017error,li2017efficient} 
and circuit knitting~\cite{bravyi2016trading,peng2020simulating,tang2021cutqc,mitarai2021constructing}. These techniques extend the size of quantum circuits that can be executed reliably on a given QPU
without resorting to error correction. We estimate the overhead introduced by state-of-the-art error mitigation methods
and  discuss recent ideas on how to combine error correction and mitigation. 
Circuit knitting techniques  exploit structural
properties of the simulated system, such as geometric locality,
to decompose a large quantum circuit into 
smaller sub-circuits or combine solutions produced by multiple QPUs.

The classical simulation algorithms used in computational physics or chemistry are often heuristics and work well in practice, even though they do not offer rigorous performance
guarantees. 
Thus, it is natural to ask whether rigorous quantum algorithms designed for simulating time evolution
admit less expensive heuristic versions that are more amenable to near-term QPUs. We discuss such algorithms in Section~\ref{sec:VQE}
to address question (2). 

To approach question (3), 
we discuss generalizations of the surface code known 
as   low-density parity check (LDPC) quantum codes~\cite{gottesman2013fault, PRXQuantum.2.040101}.
These codes can pack many more logical qubits into a given number of physical qubits such that, as the size of quantum circuits grows, only a constant fraction of physical qubits is devoted to error correction (see Section~\ref{sec:EC} for details). These more efficient codes need long-range connections between qubits embedded in a two-dimensional grid~\cite{Baspin2021nonlocality}, but the efficiency benefits are expected to outweigh the long-range connectivity costs. 

We then focus on quantum-centric supercomputing, which is a new architecture for realizing error mitigation, circuit knitting, and heuristic quantum algorithms with substantial classical calculations. At the heart of this architecture is classical and quantum integration and modularity. We need classical integration at real-time to enable conditioning quantum circuits on classical computations (dynamic circuits), at near-time to enable error mitigation and eventually error correction, and at compile time to enable circuit knitting and advanced compiling. We need modulairty to enable scaling and speeding up workflows by using parallelization.  We first start in Section \ref{sec:hardware} by focusing on superconducting computing hardware and we introduce a series of schemes—which we denote $m$, $l$, $c$, and $t$ couplers—that give us the amount of flexibility needed for realizing LDPC codes, scaling QPUs, and enabling workflows that take advantage of local operations and classical communication (LOCC) and parallelization. In Section \ref{sec:software}, we discuss the requirements on the quantum stack by defining different layers for integrating classical and quantum computations, which define requirements on latency, parallelization, and the compute instructions. From this, we can define a cluster-like architecture that we call quantum-centric supercomputer. It consists of many quantum computation nodes comprised of classical computers, control electronics, and QPUs. A quantum runtime can be executed on a quantum-centric supercomputer, working in the cloud or other classical computers to run many quantum runtimes in parallel. Here we propose that a serverless model should be used so that developers can focus on code and do not have to manage the underlying infrastructure.  We  conclude with a high level view from a developer's/user's lens.

This paper offers a perspective of the future of quantum computing focusing on an examination of what it takes to build and program near-term superconducting quantum computers and demonstrate their utility. Realizing the computational power of these machines requires the concerted efforts of engineers, physicists, computer scientists, and software developers. Hardware advances will raise the bar of quantum computers' size and fidelity. Theory and software advances will lower the bar for implementing algorithms and enable new capabilities. As both bars converge in the next few years, we will start seeing the first practical benefits of quantum computation.

\section{Towards practically useful quantum circuits}
\label{sec:useful}

Although in principle a quantum computer can reproduce any calculation performed on conventional classical hardware, the vast majority of everyday tasks are not expected to benefit from quantum-mechanical effects. However, using quantum mechanics to store and process information can lead to dramatic speedups for certain carefully selected applications. 
Of particular interest are tasks that admit a quantum algorithm with the runtime scaling as a small constant power of the problem size $n$—e.g., as $n^2$ or $n^3$—whereas the best known classical algorithm solving the problem has runtime growing faster than any constant power of $n$—e.g., as $2^n$ or $2^{\sqrt{n}}$. We define runtime as the number of elementary gates in a circuit 
(or circuits) implementing the algorithm for a given problem instance.
As the problem size $n$ grows, the more favorable scaling of the quantum runtime quickly compensates for a relatively high cost and slowness of quantum gates compared with their classical counterparts. These exponential or, formally speaking, super-polynomial speedups are fascinating from a purely theoretical standpoint and provide a compelling practical reason for advancing quantum technologies.

Known examples of tasks with an exponential quantum speedup include simulation of quantum many-body systems\cite{lloyd1996universal}, number theoretic problems such as integer factoring\cite{shor1994algorithms}, solving certain types of linear systems\cite{harrow2009quantum}, 
estimation of Betti numbers used in topological data 
analysis~\cite{lloyd2016quantum,gyurik2020towards,ubaru2021quantum},
and computing topological invariants of knots and links\cite{aharonov2009polynomial}. (We leave aside speedups obtained in the so-called Quantum RAM model\cite{giovannetti2008quantum}, for although it appears to be more powerful than the standard quantum circuit model, it is unclear whether a Quantum RAM can be efficiently implemented in  any real physical system.)

Simulation of quantum many-body systems has received the most attention due to its numerous scientific and industrial applications, and for being the original value proposition for quantum computing~\cite{Feynman82}. The ground state and thermal-equilibrium properties of many-body systems can often be understood, at least qualitatively, using classical heuristic algorithms such as dynamical mean-field theory (DMFT) or perturbative methods. However, understanding their behavior far from equilibrium in the regime governed by coherent dynamics or performing high-precision ground state simulations for strongly-interacting electrons—e.g., in the context of quantum chemistry—is a notoriously hard
problem for classical computers. 

As a simple illustration, consider a spin chain composed of $n$ quantum spins (qubits or qudits) with Hamiltonian
\[
H = \sum_{j=1}^{n-1} H_{j,j+1},
\]
where $H_{j,j+1}$ is a two-spin nearest-neighbor interaction. The Schr\"odinger equation 
\[
i\frac{d |\psi(t)\rangle}{dt} = H|\psi(t)\rangle
\]
governs the coherent time evolution of the system from some fixed initial state $|\psi(0)\rangle$. 

Suppose our goal is to compute the expected value of some local observable on the time-evolved state $|\psi(t)\rangle=e^{-iHt}|\psi(0)\rangle$. Such expected values are of great interest for understanding, among other things, thermalization mechanisms in closed quantum systems\cite{kaufman2016quantum}. Transforming the time-dependent expected values into the frequency domain provides valuable information about the excitation spectrum of the system\cite{shtanko2020unitary,aleiner2020accurately}. A slightly modified version of this problem that involves measuring each qubit of $|\psi(t)\rangle$ is known to be 
BQP-complete\cite{vollbrecht2008quantum,nagaj2008hamiltonian,kay2008computational,chase2008universal},
meaning that it is essentially as hard as simulating a universal quantum computer.  

The known classical algorithms for simulating the coherent time evolution of a quantum spin chain have runtime $\min{(2^{O(n)}, 2^{O(vt)})}$, where $v\sim \max_j \|H_{j,j+1}\|$ is the Lieb-Robinson velocity, which controls how fast information propagates through the system\cite{bravyi2006lieb}. For simplicity, we ignore factors polynomial in $n$ and $t$. The runtime $2^{O(n)}$ can be achieved using a standard state vector simulator while the runtime $2^{O(vt)}$ can be achieved by approximating $|\psi(t)\rangle$ with Matrix Product States\cite{osborne2006efficient,schollwock2011density} or by restricting the dynamics to a light cone\cite{hastings2008observations}. In general, the linear growth of the entanglement entropy with time appears to be an insurmountable obstacle for classical simulation algorithms. 

Haah, Kothari, {\it et. al.} recently found a nearly optimal quantum algorithm for  simulating the time evolution of spin chain Hamiltonians\cite{haah2021quantum} with runtime  $\tilde{O}(nt)$, where $\tilde{O}$ hides factors logarithmic in $n$, $t$, and the inverse approximation error.  This algorithm works by approximating the time evolution operator $e^{-iHt}$ by a product of simpler unitaries describing forward and backward time evolution of small blocks of spins of length $O(\log{nt})$. Assuming that the evolution time $t$ scales as a small constant power of $n$; e.g., $t\sim n$, this constitutes an exponential quantum speedup. 

\begin{figure}
  \includegraphics[width=8cm]{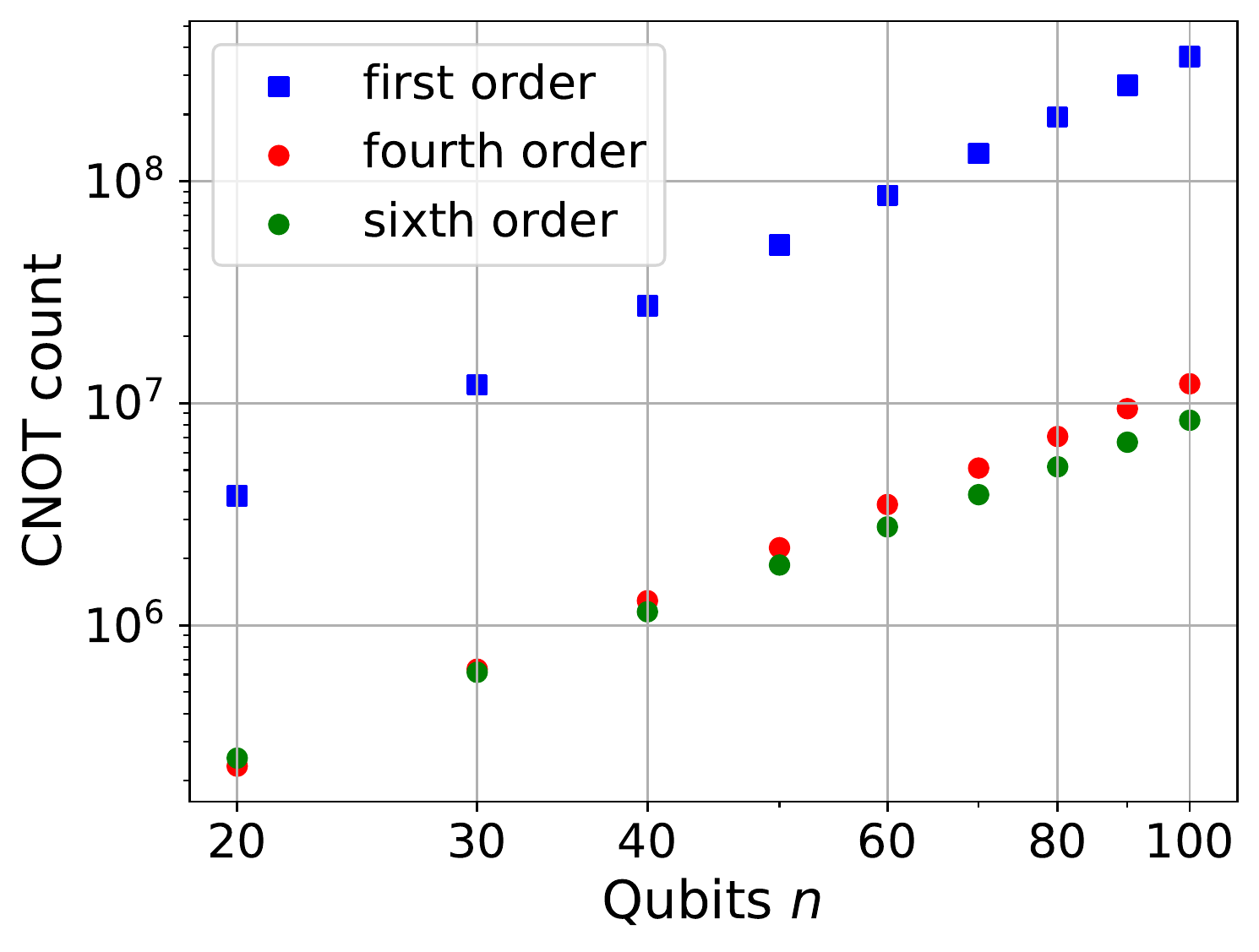}
  \caption{Estimated number of CNOT gates required to approximate the unitary evolution operator
 $e^{-iHt}$ for the $n$-qubit Heisenberg chain with $t=n$ and approximation error
$0.001$ using randomized $k$-th order product formulas ($k=1,4,6$).
The presented data is based on empirical estimates of 
ref. \onlinecite{childs2019faster},  see Eq. (70) therein, assuming that
exponentiating a single term in the Hamiltonian costs $3$ CNOTs.
 \label{fig:CNOTcount}
 }
\end{figure}

A natural question is what are the minimum quantum resources; i.e., qubit and gate counts, required to convincingly demonstrate a quantum advantage for simulating coherent dynamics. Childs, Maslov, {\it et. al.} proposed a concrete benchmark problem for this, simulating the time evolution of the spin-$1/2$ Heisenberg chain with $n=t=100$ and approximation error $0.001$\cite{childs2018toward}. The Hamiltonian has the form
$H = \sum_{j=1}^{n-1} \vec{\sigma}_j \vec{\sigma}_{j+1} + \sum_{j=1}^n h_j \sigma^z_j$, where $h_j\in [-1,1]$ are randomly chosen magnetic fields. Fig. \ref{fig:CNOTcount} shows the gate count estimates for the benchmark problem obtained by Childs, Ostrander, and Su\cite{childs2019faster}, suggesting that about $10^7$ CNOT gates (and a comparable number of single-qubit gates) are needed. This exceeds the size of quantum circuits demonstrated experimentally to date by several orders of magnitude. As we move from simple spin chain models to more practically relevant Hamiltonians, the gate count required to achieve quantum advantage
increases dramatically. For example, simulating the active space of molecules involved in catalysis problems may require about $10^{11}$ Toffoli gates\cite{berry2019qubitization}. The only viable path to reliably implementing circuits with $10^7$ gates or more  on noisy quantum hardware is quantum error correction.

\subsection{Quantum error correction}
\label{sec:EC}

One reason why conventional classical computers became ubiquitous is their ability to store and process information reliably. Small fluctuations of an electric charge or current in a microchip can be tolerated due to a highly redundant representation of logical $0$ and $1$ states by a collective state of many electrons. Quantum error-correcting codes provide a similar redundant representation of quantum states that protects them from certain types of errors. A single logical qubit can be encoded into $n$ physical qubits by specifying a pair of orthogonal $n$-qubit states $|\overline{0}\rangle$ and $|\overline{1}\rangle$ called logical-$0$ and logical-$1$. A single-qubit state $\alpha |0\rangle + \beta|1\rangle$ is encoded by the logical state $\alpha |\overline{0}\rangle+\beta|\overline{1}\rangle$. A code has distance $d$ if no operation affecting fewer than $d$ qubits can distinguish the logical states $|\overline{0}\rangle$ and $|\overline{1}\rangle$ or map them into each other. More generally, a code may have $k$ logical qubits encoded into $n$ physical qubits and the code distance $d$ quantifies how many physical qubits need to be corrupted before the logical (encoded) state is destroyed.  Thus good codes have a large distance $d$ and a large encoding rate $k/n$. 

Stabilizer-type codes\cite{gottesman1996class,calderbank1997quantum}  are by far the most  studied and promising code family. A stabilizer code is defined by a list of commuting multi-qubit Pauli observables called stabilizers such that logical states are $+1$ eigenvectors of each stabilizer. One can view stabilizers as quantum analogues of classical parity checks. Syndrome measurements aim to identify stabilizers whose eigenvalue has flipped due to errors. The eigenvalue of each stabilizer is repeatedly measured and the result—known as the error syndrome—is sent to a classical decoding algorithm. Assuming that the number of faulty qubits and gates is sufficiently small, the error syndrome provides enough information to identify the error (modulo stabilizers). The decoder can then output the operation that needs to be applied to recover the original logical state.

Most of the codes designed for quantum computing are of the LDPC type\cite{gallager1962low,tillich2013quantum,gottesman2013fault}, meaning that each stabilizer acts on only a small number of qubits and each qubit participates in a small number of stabilizers, where small means a constant independent of the code size. The main advantage of quantum LDPC codes is that the syndrome measurement can be performed with a simple constant-depth quantum circuit. This ensures that the syndrome information can be collected frequently enough to cope with the accumulation of errors. Furthermore, errors introduced by the syndrome measurement circuit itself are sufficiently benign since the circuit can propagate errors only within a ``light cone" of constant size. 

A code must satisfy several requirements to have applications in quantum computing. First, it must have a high enough error threshold—the maximum level of hardware noise that it can tolerate. If the error rate is below the threshold, the lifetime of logical qubit(s) can be made arbitrarily long by choosing a large enough code distance. Otherwise, errors can accumulate faster than the code can correct them and logical qubits can become even less reliable than the constituent physical qubits. Second, one needs a fast decoding algorithm to perform error correction in real time as the quantum computation proceeds. This may be challenging since the decoding problem for general stabilizer codes is known to be NP-hard in the worst case\cite{hsieh2011np,kuo2012hardness,iyer2015hardness}. Third, one must be able to compute on the logical qubits without compromising the protection offered by the code. In the sub-threshold regime, one must be able to realize arbitrarily 
precise logical gates from some universal gate set by choosing a large enough code distance.

The 2D surface code\cite{kitaev2003fault,bravyi1998quantum} has so far been considered an uncontested leader in terms of the error threshold—close to $1\%$ for the commonly studied depolarizing noise\cite{raussendorf2007fault,fowler2009high,wang2011surface}—yet has two important shortcomings. First, allocating a roughly $d\times d$ patch of physical qubits for each logical qubit incurs a large overhead. Unfortunately, it was shown\cite{bravyi2010tradeoffs} that any 2D stabilizer code has encoding rate $k/n=O(1/d^2)$ which vanishes for large code distance. This means that as one increases the degree of protection offered by the surface code, quantified by the code distance $d$, its encoding rate approaches zero. That is, as the size of quantum circuits grows, the vast majority of physical qubits are devoted to error correction. This is a known fundamental limitation of all quantum codes that can be realized locally in the 2D geometry. 

To make error correction more practical and minimize qubit overhead, codes with a large encoding rate $k/n$ are preferable. For example, quantum LDPC codes can achieve a constant encoding rate independent of the code size\cite{tillich2013quantum}. In fact, the encoding rate can be arbitrarily close to one\cite{gottesman2013fault}. 
A recent breakthrough result~\cite{panteleev2021asymptotically} demonstrated the existence of so-called good quantum LDPC codes that combine a constant encoding rate $k/n$ (which can be 
arbitrarily close to $1$) and a linear distance $d\ge cn$ for some constant $c>0$. For comparison, the 2D surface code has an asymptotically vanishing encoding rate and has distance at most $\sqrt{n}$.
Certain LDPC codes have a favorable property known as single-shot error correction\cite{bombin2015single,kubica2021single}. They provide a highly redundant set of low-weight Pauli observables (known as gauge operators) that can be measured to obtain the error syndrome more efficiently. This reduces the number of syndrome measurement cycles per logical gate from $O(d)$ to $O(1)$ and hence enables very fast logical gates. The syndrome measurement circuit for a quantum LDPC code requires a qubit connectivity dictated by the structure of stabilizers, i.e., one must be able to couple qubits that participate in the same stabilizer. Known examples of LDPC codes with a single-shot error correction require 3D or 4D geometry\cite{bombin2015single,kubica2021single,campbell2019theory}.

The second shortcoming of the surface code is the difficulty of implementing a computationally universal set of logical gates~\cite{bravyi2013classification}.
The surface code and its variations such as the honeycomb code~\cite{hastings2021dynamically} or folded surface code~\cite{moussa2016transversal} offer a low-overhead implementation of logical Clifford gates such as CNOT, Hadamard $H$, and phase shift $S$. These gates can be realized by altering the pattern of stabilizers measured at each time step using the code deformation method. 
However, Clifford gates are not computationally universal on their own. A common strategy for achieving universality is based on preparation of logical ancillary states $(|\overline{0}\rangle + e^{i\pi/4} |\overline{1}\rangle)/\sqrt{2}$ known as magic states. A magic state is equivalent (modulo Clifford operations) to a single-qubit gate $T=\mathrm{diag}(1,e^{i\pi/4})$.  The Clifford+$T$ gate set is universal and has a rich algebraic structure enabling efficient and nearly-optimal compiling of quantum algorithms\cite{kliuchnikov2012fast,ross2014optimal}. Unfortunately, the overhead for distilling high-fidelity magic states is prohibitively large. O’Gorman and Campbell\cite{o2017quantum} performed a careful examination of available distillation methods and their overhead, considering the implementation of a logical Clifford+$T$ circuit of size $N$ with an overall fidelity of $90\%$. Assuming a physical error rate  of $10^{-3}$, the following fitting formulas were found for the space-time volume ((physical qubits) $\times$ (syndrome measurement cycles)) associated with a single logical gate:
\begin{center}
\begin{tabular}{c|c}
Logical gate & Physical space-time volume  \\
\hline
\hline
CNOT & $1610+45\left( \log_{10}{N}\right)^{2.77}$ \\
\hline
$T$-gate & $3.13+3220 \left( \log_{10}{N}\right)^{3.20}$ \\
\end{tabular}
\end{center}
The space-time volume roughly quantifies the number of physical gates required to implement a single logical gate. 

As an example, consider the Heisenberg benchmark problem 
described above with $100$ logical qubits. The desired time evolution operator 
can be approximated
using about $10^7$ CNOTs and single-qubit gates (see fig.~\ref{fig:CNOTcount}). However, each single-qubit gate needs to be compiled using the logical-level gate set $\{H,S,T\}$. In total, this  requires roughly  $10^{9}$ $T$-gates
and a comparable number of Clifford gates\cite{childs2018toward}. Accordingly, the physical space-time volumes of a single logical CNOT and $T$-gate are roughly $2\times10^4$ and $4\times 10^6$, respectively. (In fact, this underestimates the error correction overhead
since the Heisenberg benchmark problem
requires logical circuit fidelity $0.999$
rather than $0.9$, as considered in ref. \onlinecite{o2017quantum}.)

The large overhead associated with logical  non-Clifford gates may rule out the near-term implementation of error-corrected quantum circuits, even if fully functioning logical qubits based on the surface code become available soon. There have been several strategies proposed recently for reducing this overhead, including high-yield magic state distillation methods\cite{haah2017magic,hastings2018distillation}, better strategies for preparing ``raw" noisy magic states that reduce the required number of distillation rounds\cite{li2015magic}, and better surface code implementations of distillation circuits
\cite{fowler2012bridge,litinski2019game,paetznick2013quantum,litinski2019magic}. A recent breakthrough result by
Benjamin Brown\cite{brown2020fault} 
showed how to realize a logical non-Clifford gate CCZ (controlled-controlled-$Z$) in the 2D surface code architecture without resorting to state distillation. This approach relies on the fact that a 3D version of the surface code enables an easy (transversal) implementation of a logical CCZ~\cite{kubica2015unfolding,vasmer2019three}  and a clever embedding of the 3D surface code into a 2+1 dimensional space-time. It remains to be seen whether this method is competitive compared with magic state distillation.

\subsection{Error mitigation}
\label{sec:EM}

Although error correction is vital for realizing large-scale quantum algorithms with great computational power, it may be overkill for small or medium size computations. A limited form of correction for shallow quantum circuits can be achieved by combining the outcomes of multiple noisy quantum experiments in a way that cancels the contribution of noise to the quantity of interest\cite{temme2017error,li2017efficient}. These methods, collectively known as error mitigation, 
are well suited for the QPUs available today because they introduce little to no overhead in terms of the number of qubits and only a minor overhead in terms of extra gates. However, error mitigation comes at the cost of an increased number of circuits (experiments) that need to be executed. In general, this will result in an exponential overhead; however, the base of the exponent can be made close to one with improvements in hardware and control methods, and each experiment can be run in parallel. Furthermore, known error mitigation methods apply only to a restricted class of quantum algorithms that use the output state of a quantum circuit to estimate the expected value of observables. 

Probabilistic error cancellation (PEC) \cite{temme2017error,endo2018practical} aims to approximate an ideal quantum circuit via a weighted sum of noisy circuits that can be implemented on a given quantum computer. The weights assigned to each noisy circuit can be computed analytically if the noise in the system is sufficiently well characterized or learned by mitigating errors on a training set of circuits that can be efficiently simulated classically\cite{strikis2020learning}. We expect that the adoption of PEC will grow due to the recent theoretical and experimental advances in quantum noise metrology~\cite{harper2020efficient,flammia2021averaged,vandenBerg2022pec}. For example, ref. \onlinecite{vandenBerg2022pec} shows how to model the action of noise associated with a single layer of two-qubit gates by a Markovian dynamics with correlated Pauli errors.  This model can be described by 
a collection of single-qubit and two-qubit Pauli errors $P_1,\ldots,P_m$ and the associated error rate
parameters $\lambda_1,\ldots,\lambda_m\ge 0$ such that the combined noise channel
acting on a quantum register has the form $\Lambda(\rho)=\exp{[{\cal L}]}(\rho)$, where
$\cal L$ is a Lindblad generator, ${\cal L}(\rho)  =\sum_{i=1}^m \lambda_i (P_i \rho P_i^\dag - \rho)$.
The unknown error rates $\lambda_i$ can be learned to 
within several digits of precision by repeating
the chosen layer of gates many times and measuring the decay
of suitable observables~\cite{vandenBerg2022pec}.
The error mitigation overhead (as measured by the number of circuit repetitions) per layer of gates
scales as $\gamma^2$, where 
\[
\gamma=\exp{\left(2\sum_{i=1}^m  \lambda_i\right)}.
\]

For a circuit composed of $d>1$ layers, the error rates $\lambda_i$ may be layer-dependent
and have to be learned separately for each layer. As observed in ref. \onlinecite{vandenBerg2022pec}, this model
can approximate the actual hardware noise very well 
using only $m=O(n)$ elementary Pauli errors $P_i$
supported on edges of the qubit connectivity graph,
where $n$ is the total number of qubits in the circuit. In general, the runtime for getting a noise-free estimate will depend on the circuit implemented and the noise model used. 

A large class of quantum algorithms
that can benefit from PEC
is based on the so-called hardware-efficient circuits\cite{Kandala_2017}.
A depth-$d$ hardware-efficient circuit
consists of $d$ layers of  two-qubit gates such that all gates within the same layer are non-overlapping and couple qubits that are nearest-neighbors in the QPU connectivity graph.
 Denoting the average error rate
per qubit $\bar{\lambda} = (1/n)\sum_{i=1}^m \lambda_i$,
averaged over all $d$ layers, 
the overall PEC overhead scales as  
$(\bar{\gamma})^{dn}$,
where $\bar{\gamma}=\exp{(4\bar{\lambda})}$. This allows a simple formula for estimating the runtime, $J$, for a noise-free estimate from a quantum circuit of depth $d$ and width $n$ to be 
\begin{equation}
\mathrm{J} =d (\bar{\gamma})^{dn} \beta ,
\label{gammabeta}
\end{equation}
where $\beta$ is the average time to run a single layer of the circuit. One can view $\beta$ as a measure of the ``speed" and $\bar{\gamma}$ as a hardware-dependent parameter
that quantifies the average ``quality" of gates across the entire QPU.

\begin{figure}
 \centerline{
    \includegraphics[width=8cm]{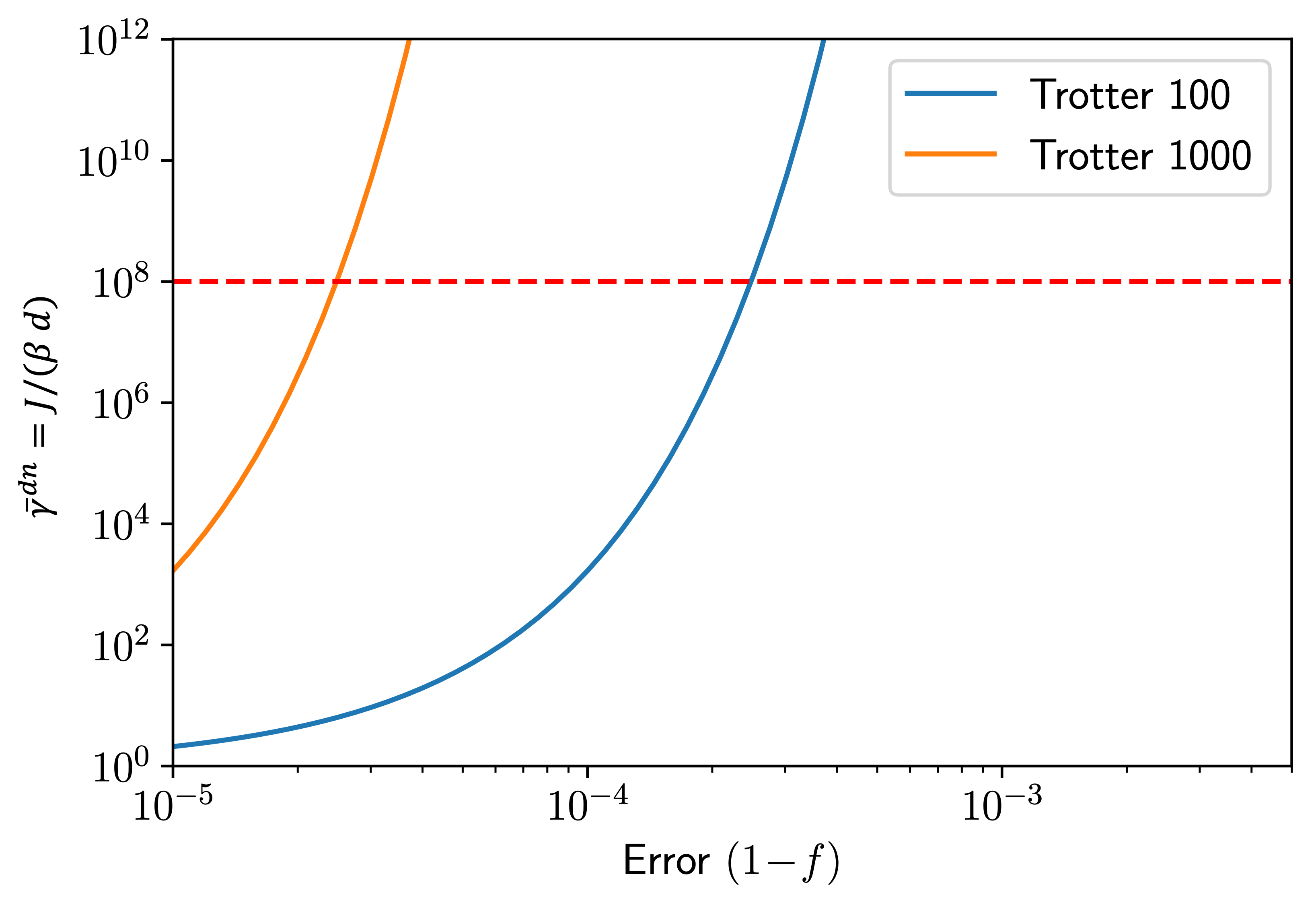}}
  \caption{Runtime scaling (number of circuit instances) needed
to error mitigate 100 and 1000 Trotter steps in a circuit of 100 qubits and layers of non-overlapping two-qubit gates, each gate affected by a local depolarizing two-qubit error. The red dotted line identifies 100 million circuits, the daily limit assuming a repetition rate of 1 $ ms$. So far the only architectures that have achieved this speed are solid-state based. The number of circuit instances dramatically decreases with slight improvements in the error rate of the physical gates.
 \label{fig:EMscaling}
 }
\end{figure}

For a spin chain of $n = 100$ qubits, the size of our benchmark problem, fig. ~\ref{fig:EMscaling} shows the number of circuit instances that need to be sampled to perform PEC for 100 and 1000 Trotter steps using the decomposition in fig. 3A of ref. \onlinecite{vandenBerg2022pec}. Current hardware runs of up to $10^8$ circuits daily (red dashed line) and error rates of $10^{-3}$ have been demonstrated. Hence, we anticipate that with a couple orders of magnitude improvement this becomes possible. Furthermore, this runtime can be further reduced with the quantum-centric supercomputing architecture that allows parallelized execution of quantum circuits.

We can also measure the quantity of interest at several different values of the noise rate and perform an extrapolation to the zero-noise limit~\cite{temme2017error,li2017efficient,kandala2019error}. This method cancels the leading-order noise contribution
as long as the noise is weak and Markovian. Unlike PEC, this method is biased and heuristic but may require fewer circuits for the reconstruction. This method was recently demonstrated\cite{kim2021} to scale up to 27 qubits and still reconstruct observables. Whether this method can be combined with PEC, which gives an unbiased estimation, remains an open question.

More general (non-Markovian) noise can be mitigated using the virtual distillation technique\cite{huggins2020virtual,koczor2020exponential}. It works by combining two copies of a noisy output state $\rho$ in a way that enables measurements of observables on a  state $\rho^2/\mathrm{Tr}(\rho^2)$. Assuming that $\rho$ has most of its weight on the ideal output state, virtual distillation can quadratically suppress the contributions of errors. However, this method introduces at least a factor of two overhead in the number of qubits and gates. A review of the existing error mitigation proposals can be found in Endo, {\it et. al.}\cite{endo2021hybrid}.

We anticipate error mitigation to continue to be relevant when error-corrected QPUs with a hundred or more logical qubits become available. As discussed in Section \ref{sec:EC}, the first generation of error-corrected quantum chips based on 2D stabilizer codes may not be able to execute universal computations. Such QPUs are likely to offer only high-fidelity Clifford gates such as the Hadamard or CNOT, which can all be
efficiently simulated on a classical computer.
Meanwhile, logical non-Clifford gates such as the $T$-gate may remain out of reach due to the need to perform magic state distillation. This leads to the interesting possibility of combining error correction and mitigation. A concrete proposal by Piveteau, {\it et al.}\cite{piveteau2021error} leverages the ability to realize {\em noisy} logical $T$-gates with fidelity comparable to or exceeding that of physical (unencoded) gates. Applying error mitigation protocols at the logical level to cancel errors introduced by noisy $T$-gates enables one to simulate universal logical circuits without resorting to state distillation. This may considerably reduce the hardware requirements for achieving a quantum advantage. However, error mitigation comes at the cost of an increased number of circuit executions. Assuming a physical gate fidelity of $99.9\%$ and a budget of $1,000$ circuit executions, Piveteau, {\it et al.}\cite{piveteau2021error} estimate that logical Clifford+$T$ circuits with about 2,000 $T$-gates can be realized reliably. This is far beyond the limit of existing classical algorithms that can simulate Clifford+$T$ circuits with about 50 $T$-gates~\cite{bravyi2016improved,bravyi2019simulation}. Similar ideas for combining error correction and mitigation are discussed in refs. \onlinecite{lostaglio2021error,suzuki2020quantum}.

\subsection{Circuit Knitting}
\label{sec:CK}

We can extend the scope of near-term hardware to compensate for other shortcomings such as a limited number of qubits or qubit connectivity by using circuit knitting techniques. This refers to the process of simulating small quantum circuits on a quantum computer and stitching their results into an estimation of the outcome of a larger quantum circuit. As was the case with error mitigation, known circuit knitting techniques apply to a restricted class of quantum algorithms that aim to
estimate the expected value of observables.

The most well-known example is circuit cutting \cite{bravyi2016trading,peng2020simulating,tang2021cutqc,mitarai2021constructing}. 
In this method, a large quantum circuit is approximated by a weighted sum of circuits consisting of small isolated sub-circuits. Each sub-circuit can be executed separately on a small QPU. 
The overhead introduced by this method (as measured by the number of circuit repetitions) 
scales exponentially with the number of two-qubit gates or qubit wires that need to be cut in order to achieve the desired partition of the circuit. Surprisingly, it was recently shown that the circuit cutting overhead can be substantially reduced by running the isolated sub-circuits in parallel using non-interacting QPUs that can only exchange classical data\cite{piveteau2022circuit}. This approach requires hardware capable of 
implementing dynamic circuits\footnote{Dynamic circuits are computational circuits that combine quantum and classical operations, using the outcome of classical computations to adapt subsequent quantum operations. For more information, see section \ref{sec:software}.}, where the control electronics is extended to include independent QPUs.

A second example is entanglement forging\cite{eddins2021doubling}, where either an entangled variational state is decomposed into a weighted sum of product states or the entanglement between a pair of qubit registers is converted into time-like correlations within a single register\cite{huembeli2022entanglement}. The overhead of this method  typically scales exponentially with the 
amount of entanglement across the chosen partition of the system. 

A third example, closely related to circuit knitting, 
uses embedding methods to
decompose the simulation of a 
large quantum many-body system into smaller subsystems that can be simulated
individually on a QPU. The interactions between subsystems are
accounted for by introducing an effective bath that could be either a classical
environment or another small quantum system.
The decomposition of the original system and the optimization of the bath parameters are
performed on a classical computer that can exchange classical data with the
QPU. Well-known examples of quantum embedding methods
that build on their classical counterparts are dynamical mean-field theory~\cite{Bauer15,kreula2016few,bravyi2017complexity},
density-matrix embedding~\cite{knizia2012density,knizia2013density,mineh2022solving}, and density-functional embedding~\cite{ma2020quantum}.

\subsection{Heuristic quantum algorithms}
\label{sec:VQE}

Heuristic quantum algorithms can be employed near-term to solve classical optimization~\cite{FGG2014}, machine learning~\cite{havlivcek2019supervised}, and quantum simulation~\cite{yuan2019theory} problems. These fall into two categories—algorithms that use kernel methods\cite{havlivcek2019supervised} and variational quantum algorithms (VQA). Quantum kernel methods have also been found that lead to provable speedups\cite{liu2020shor} and expand to a class of kernels for data with group structure\cite{glick2021covariant}. For VQA, the basic proposal is appealingly simple: an experimentally-controlled trial state is used as variational wavefunction to minimize the expected energy of a given quantum Hamiltonian or a classical cost function encoding the problem of interest. The trial state is usually defined as the output state of a shallow quantum circuit.  Rotation angles that define individual gates serve as variational parameters. These parameters are adjusted via a classical feedback loop to optimize the chosen cost function.

At present, there is no mathematical proof that VQA can outperform classical algorithms in any task. In fact, it is known that VQA based on sufficiently shallow (constant depth) variational circuits with 2D or 3D qubit connectivity can be efficiently simulated on a classical computer~\cite{bravyi2021classical,coble2020quasi}. This rules out a quantum advantage. Meanwhile, the performance of VQA based on deep variational circuits is severely degraded by noise~\cite{franca2020limitations}. However, as the error rates of QPUs decrease, we should be able to execute VQA in the intermediate regime where quantum circuits are already hard to simulate classically but the effect of noise can still be mitigated. 

As a concrete example, let us discuss possible applications of VQA to the problem of simulating coherent time evolution of quantum spin chains. It is commonly believed~\cite{haah2021quantum} that approximating the time evolution operator $e^{-iHt}$ for a Hamiltonian $H$ describing a 1D chain of $n$ qubits requires a quantum circuit of size scaling at least linearly with the space-time volume $nt$. Meanwhile, the best known rigorous quantum algorithms based on product formulas~\cite{childs2019faster} or Lieb-Robinson bounds~\cite{haah2021quantum} require circuits of size  $O(n^2 t)$ or $O(nt \cdot \mathrm{polylog}(nt))$, respectively. A natural question is whether VQA can reduce the circuit size to what is believed to be optimal, that is, linear in $nt$. If this is indeed the case, a QPU with gate fidelity around $99.99\%$ 
may be able to solve the classically hard problem instances described
above with space-time volume  $nt\sim 10^4$ using existing error mitigation techniques.
 
Variational quantum time evolution (VarQTE) algorithms, pioneered by Li and Benjamin~\cite{li2017efficient}, could be an alternative to simulate the time evolution  of  these classically hard instances given near-term noisy QPUs. These algorithms aim to approximate the time-evolved state $|\psi(t)\rangle = e^{-iHt}|\psi(0)\rangle$ by a time-dependent variational ansatz $|\phi(\theta)\rangle=U(\theta)|0^n\rangle$,
where $U(\theta)$ is a parameterized  quantum circuit  with a fixed layout of gates and $\theta=\theta(t)$ is a vector of variational parameters. The initial state $|\psi(0)\rangle$ is assumed to be sufficiently simple so that the variational ansatz 
for $|\psi(0)\rangle$ is easy to find. The goal is to find a function $\theta(t)$ such that the variational state $|\phi(\theta(t))\rangle$ approximates the time evolved state $|\psi(t)\rangle$ for  all $t$ in the chosen time interval. As shown in ref. \onlinecite{li2017efficient}, the desired function $\theta(t)$ can be efficiently computed using the stationary-action principle with a Lagrangian $L(t)=\langle \phi(\theta(t))|d/dt + iH|\phi(\theta(t))\rangle$. This yields a first-order differential equation~\cite{yuan2019theory}
$\sum_q M_{p,q} \dot{\theta}_q= V_p$ where
\[
M_{p,q} =  \mathrm{Im}( \langle \partial_p \phi(\theta)|\partial_q \phi(\theta)\rangle),
\]
\[
V_p = -\mathrm{Re}(\langle \partial_p \phi(\theta)|H|\phi(\theta)\rangle),
\]
and $\partial_p \equiv \frac{\partial}{\partial \theta_p}$.
As shown in [\onlinecite{li2017efficient}], the entries of $M$ and $V$ can be efficiently estimated on a quantum computer. A comprehensive review of VarQTE algorithms can be found in refs. \onlinecite{yuan2019theory,barison2021efficient}.

The fact that VarQTE algorithms are heuristics and therefore
 lack rigorous
performance guarantees raises the question of how to validate them. This becomes particularly important for large problem sizes
where verifying a solution of the problem on a classical computer becomes impractical. Ref. \onlinecite{zoufal2021error} recently developed a version of VarQTE based on McLachlan’s variational principle that comes with efficiently computable bounds on the distance between the exact time-evolved state $|\psi(t)\rangle$ and the approximate variational state found by the VarQTE algorithms. Thus, although VarQTE lacks a rigorous justification,
one may be able to obtain {\em a posteriori} bounds on its approximation error for some specific problems of practical interest. 

\subsection{Summary}

To summarize, the Heisenberg chain example illustrates what we believe are general guidelines for designing near-term quantum algorithms. 

First, our best chance of attaining a quantum advantage is by focusing on problems that admit an exponential (super-polynomial) quantum speedup. Even though a quantum algorithm that achieves such speedup with formal proof may be out of reach for near-term hardware, its mere existence serves as  compelling evidence that quantum-mechanical effects such as interference or entanglement are beneficial for solving the chosen problem. 

Second, the only known way to realize large-scale quantum algorithms relies on quantum error-correcting codes. The existing techniques based on the surface code are not satisfactory due their poor encoding rate
and
high cost of logical non-Clifford gates. Addressing these shortcomings may require advances in quantum coding theory such as developing high-threshold fault-tolerant protocols based on quantum LDPC codes and improving the qubit connectivity of QPUs beyond the 2D lattice. Supplementing error correction with cheaper alternatives such as error mitigation and circuit knitting may provide a more scalable way of implementing high-fidelity quantum circuits. 

Third, near-term quantum advantage should be possible by exploring less expensive, possibly heuristic versions of the algorithm considered. Those heuristic quantum algorithms lack rigorous performance guarantees, but they may be able to certify the quality of a solution {\em a posteriori} and offer a way to tackle problems that cannot be simulated classically. 

We believe these general guidelines define the future of quantum computing theory and will guide us to important demonstrations of its benefits for the solution of scientifically important problems in the next few years.

\section{The Path to Large Quantum Systems}
\label{sec:hardware}

The perspective above leads to a challenge in quantum hardware.  We believe there will be near-term advantage using a mixture of error mitigation, circuit knitting and heuristic algorithms. On a longer time frame, partially error-corrected systems will become critical to running more advanced applications and further down the line, fault-tolerant systems running on not-as-yet fully explored LDPC codes with non-local checks will be key.  The first steps for all of these approaches are the same: we need hardware with more qubits capable of higher fidelity operations.  We need tight integration of fast classical computation to handle the high run-rates of circuits needed for error mitigation and circuit knitting, and the classical overhead of the error correction algorithm afterwards. This drives us to identify a hardware path that starts with the early heuristic small quantum circuits and grows until reaching an error-corrected computer.

\subsection{Cycles of Learning}

The first step in this path is to build systems able to demonstrate near-term advantage with error mitigation and limited forms of error correction.
Just a few years ago, QPU sizes were limited by control electronics cost and availability, I/O space, quality of control software, and a problem referred to as ``breaking the plane"\cite{Gambetta2017}, i.e., routing microwave control and readout lines to qubits in the center of dense arrays. Today, solutions to these direct barriers to scaling have been demonstrated, which has allowed us to lift qubit counts beyond 100—above the threshold where quantum systems become intractably difficult to simulate classically and examples of quantum advantage become possible.  The next major milestones are (1) increasing the fidelity of QPUs enough to allow exploration of quantum circuits for near-term quantum advantage with limited error correction and (2) improving qubit connectivity beyond 2D—either through modified gates, sparse connections with non-trivial topologies, and/or increasing the number of layers for quantum signals in 3D integration—to enable the longer term exploration of efficient non-2D LDPC error-correction codes. These developments are both required for our longer term vision, but can be pursued in parallel.

\begin{figure*}
\centerline{
    \includegraphics[width=0.75\textwidth]{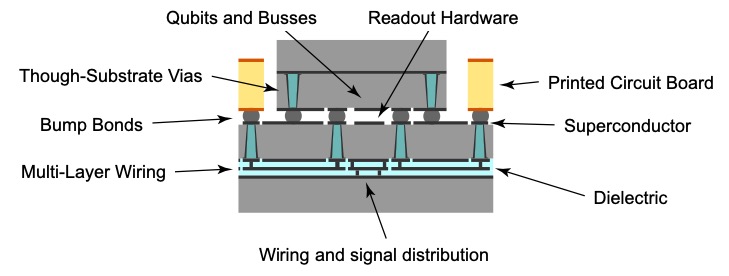}}
    \caption{An example of a scheme that allows breaking the plane for signal delivery compatible with the integration of hundreds of qubits. It is composed of technologies adapted from conventional CMOS processing.\label{fig:plane_break}}
\end{figure*}

Work on improving the quality of quantum systems by improving gate fidelities involves many cycles of learning, trying coupling schemes, process changes, and innovations in controlling coupling and crosstalk.  Scaling this work to large QPUs capable of demonstrating quantum advantage, and ultimately to the extreme system scales we anticipate in the distant future, involves integrating different technologies with enough reliability and skill to make size be limited by cost and need, not by technological capability. This adds challenges in reliability, predictability, and manufacturability of QPUs while continuing to incorporate improved technologies into these complex systems.  Meanwhile, the increased development, fabrication, and test times for larger systems creates a lag in cycles of innovation that must be overcome.

The manufacturing cycle time increases with QPU sophistication. Many simple transmon QPUs require just a single level of lithography and can be easily fabricated in a day or two. Even the 5- and 16-qubit QPUs that were IBM's initial external cloud quantum systems involved only two lithography steps and took a week to fabricate. Compare this to more advanced packaging schemes like those at MIT Lincoln Laboratory\cite{Yost2020, Tolpygo2015, Rosenberg2019} or IBM's newer ``Eagle'' QPUs (fig. \ref{fig:plane_break}), which involve dozens of lithography steps and slow process steps, and take months to build at a research-style facility with one-of-a-kind tools. This increased cycle time makes it harder to reach the fidelities and coherence times needed as well as debug the manufacturing and assembly for reliable QPU yield.
 
Reliability in semiconductor manufacturing is not a new problem.  In general, among the unique component challenges faced in building a scaled machine, the conventional semiconductor technologies integrated on chip are the most well studied. Incorporating them in superconducting technologies is more a matter of ensuring that the associated processes are compatible with each other than inventing new approaches. However, the rapid growth of volume we anticipate being needed is a major challenge.

Many failure modes in superconducting quantum systems are not detectable until the QPUs are cooled to their operating temperature, sub-100 mK.  This is a severe bottleneck that renders in-line test (where a device sub-component is tested for key metrics before the QPU build finishes) and process feed-forward (where future process steps are modified to correct for small deviations in early steps and stabilize total device performance) difficult or impossible. There are exceptions where it is possible to tightly correlate an easy measurement at room temperature with ultimate QPU performance: for example, resistance measurements of Josephson junctions can accurately predict their critical currents and hence, the frequency of qubits made with them—a key parameter in fixed frequency systems. We can take advantage of these statistical correlations wherever they exist for rapid progress in parts of our process \cite{Kreikebaum2020} or in post-process tuning \cite{2009.00781}.  However, reliably establishing these correlations requires measuring hundreds or thousands of devices, a nontrivial feat.

Absent these correlations, we can use simplified test vehicles; for example, rather than using the entire complicated signal delivery stack when trying to improve qubit coherence, we can use a simplified device designed to obtain good statistics and fast processing \cite{7745914}.  Still, identifying specific steps leading to increased coherence is nontrivial. It is rarely possible to change just one parameter in materials processing. Changing a metal in a qubit may also change etch parameters, chemicals compatible with the metal for subsequent processing, and even allowed temperature ranges\cite{Place2021}. Once an improved process is found,
it is hard to identify exactly which steps were critical vs. simply expedient.  

We must gather sufficient statistics when performing materials  research for the results be meaningful and provide enough certainty\cite{McRae2021}. We should carefully document process splits wherever relevant, and we should publish changes in materials processes that lead to neutral or even negative results, not just just publish highly successful work.

Similar difficulties occur in non-material based research on devices. Some gates work well between pairs of qubits yet exhibit strong couplings that make them unsuitable for larger QPUs or compromise single-qubit performance.  Three- and four-qubit experiments are no longer challenging from a technical or budgetary perspective. To be relevant to larger QPUs, research needs to move away from two-qubit demos, especially hero experiments between a single pair of qubits in which many critical defects can be masked by luck.  

A mixture of long cycle-time complex devices and short cycle-time test vehicles for sub-process development and quantum operations is key to continuing improvements in the quality of QPUs and provides a recipe for continued R\&D contributions as the largest QPUs begin to exceed the capabilities of smaller groups and labs. Nonetheless, reductions in long cycle times are needed.  Some of this will come naturally—first-of-a-kind processes and QPUs usually take longer as they tend to include extra steps, inspections, and in-line tests that, while suggested by general best practices, may not be necessary.  While counterproductive from a cost viewpoint, building the ``same'' QPU repeatedly to iron out manufacturing problems and speed up cycles of innovation will likely be a successful strategy for the largest QPUs with the most complex fabrication flows.

\subsection{Supporting Hardware}

Scaling to larger systems also involves scaling classical control hardware and the input/output (I/O) chain in and out of the cryostat. This I/O chain, while still needing substantial customization for the exact QPU being controlled, consists of high volumes of somewhat more conventional devices; for example, isolators, amplifiers, scaled signal delivery systems, and more exotic replacements such as non-ferrite isolators and quantum limited amplifiers that may offer performance, cost, or size improvements.  These components have enormous potential for being shared between various groups pursuing quantum computing, and in some instances can be purchased commercially already.  However, assembling these systems at the scale required today, let alone a few years time, requires a high volume cryogenic test capability that does not currently exist in the quantum ecosystem, creating a short-term need for vertically-integrated manufacturing of quantum systems. The challenge here is establishing a vendor and test ecosystem capable of scaled, low-cost production—a challenge made difficult by the fact that the demand is somewhat speculative.

There are also one-off components per system; for example, each quantum computer we deploy only requires a single dilution refrigerator, or in many cases a fraction thereof. The dilution refrigerator manufacturer effectively acts as a systems integrator for cryo-coolers, wiring solutions, pumping systems, and even some active electronics. Maintaining the flexibility we need to change quickly as the systems scale will be most easily attainable if we can standardize many of these interfaces so that, for example, moving to a more scalable cooling technology at 4K doesn't require redesigning the entire refrigeration infrastructure.

Currently, each group building large QPUs has their own bespoke control hardware. Given the radically different control paradigms and requirements\cite{Geck2019,Xue2021,Ryan2017,Pino2021}, it is unlikely that the analog front-ends of these systems could ever be shared.  However, there is a common need for sequencing logic (branching, local and non-local conditionals, looping) at low-cost and low-power for all types of quantum computers, not just solid-state. 
These will likely need to be built into a custom processor—an Application Specific Integrated Circuit or ASIC—as we scale to thousands of qubits and beyond. On top of this, the software that translates a quantum circuit into the low-level representation of this control hardware is becoming increasingly complex and expensive to produce. Reducing cost favors a common control platform with customized analog front ends. Open-specification control protocols like OpenQASM3\cite{qasm3} are already paving the way for this transformation.

\subsection{Classical parallelization of quantum processors}\label{clapar}

Reaching near-term quantum advantage will require taking advantage of techniques like circuit knitting and error mitigation that effectively stretch the capabilities of QPUs—trading off additional circuit executions to emulate more qubits or higher fidelities.  These problems can be pleasingly parallel, where individual circuits can execute totally independently on multiple QPUs, or may benefit from the ability to perform classical communication between these circuits that span multiple QPUs.  Introduction of control hardware that is able to run multiple QPUs as if they were a single QPU with shared classical logic, or split a single QPU into multiple virtual QPUs to allow classical parallelization of quantum workloads is an important near-term technology for stretching this advantage to the limit.  Longer term, these technologies will play a critical enabling role as we begin to build quantum systems that span mutliple chips and multiple cryostats, i.e., modular quantum systems.

\subsection{Modularity}

\begin{figure*}
\centering
    \subfigure[{\it p} type modularity for classical parallelization of QPUs]{\includegraphics[width=0.49\textwidth]{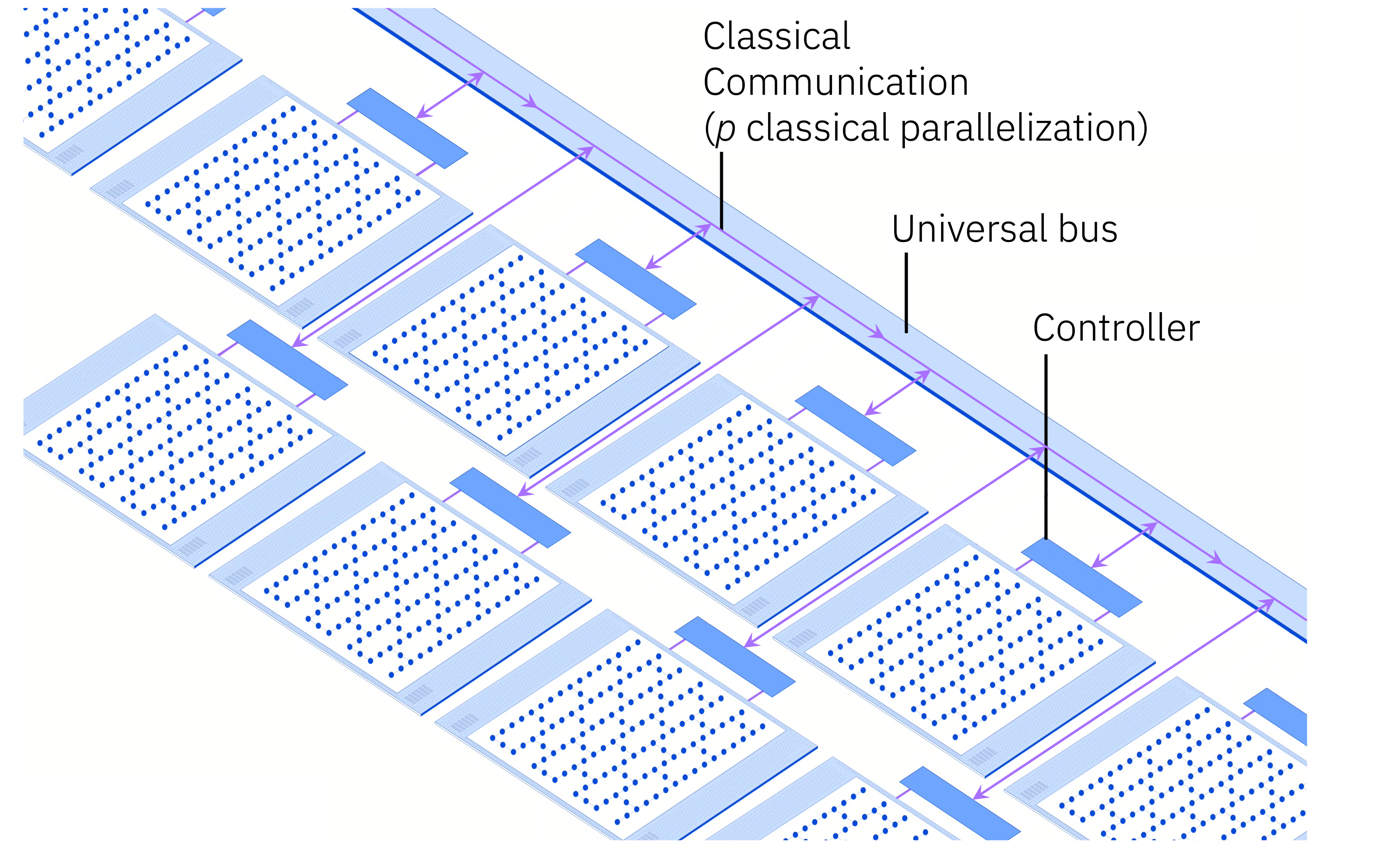}}
    \subfigure[Dense modularity {\it m} and on-chip non-local couplers {\it c} for LDPC codes for creating a single QPU from multiple chips ]{\includegraphics[width=0.49\textwidth]{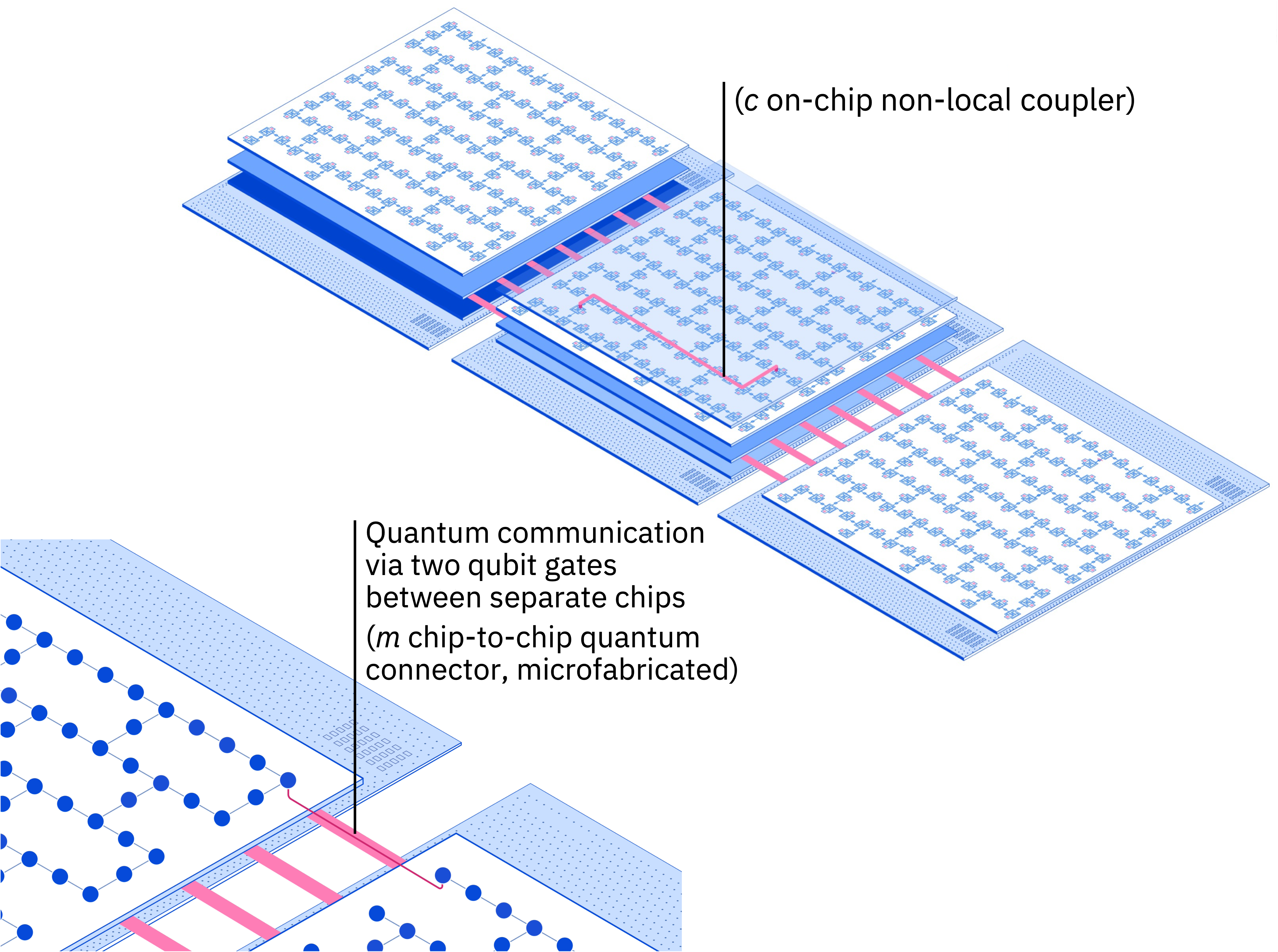}} 
    \\ \vspace{0.3cm}
    \subfigure[Long-range {\it l} type modularity to enable quantum parallelization of multiple QPUs ]{\includegraphics[width=0.49\textwidth]{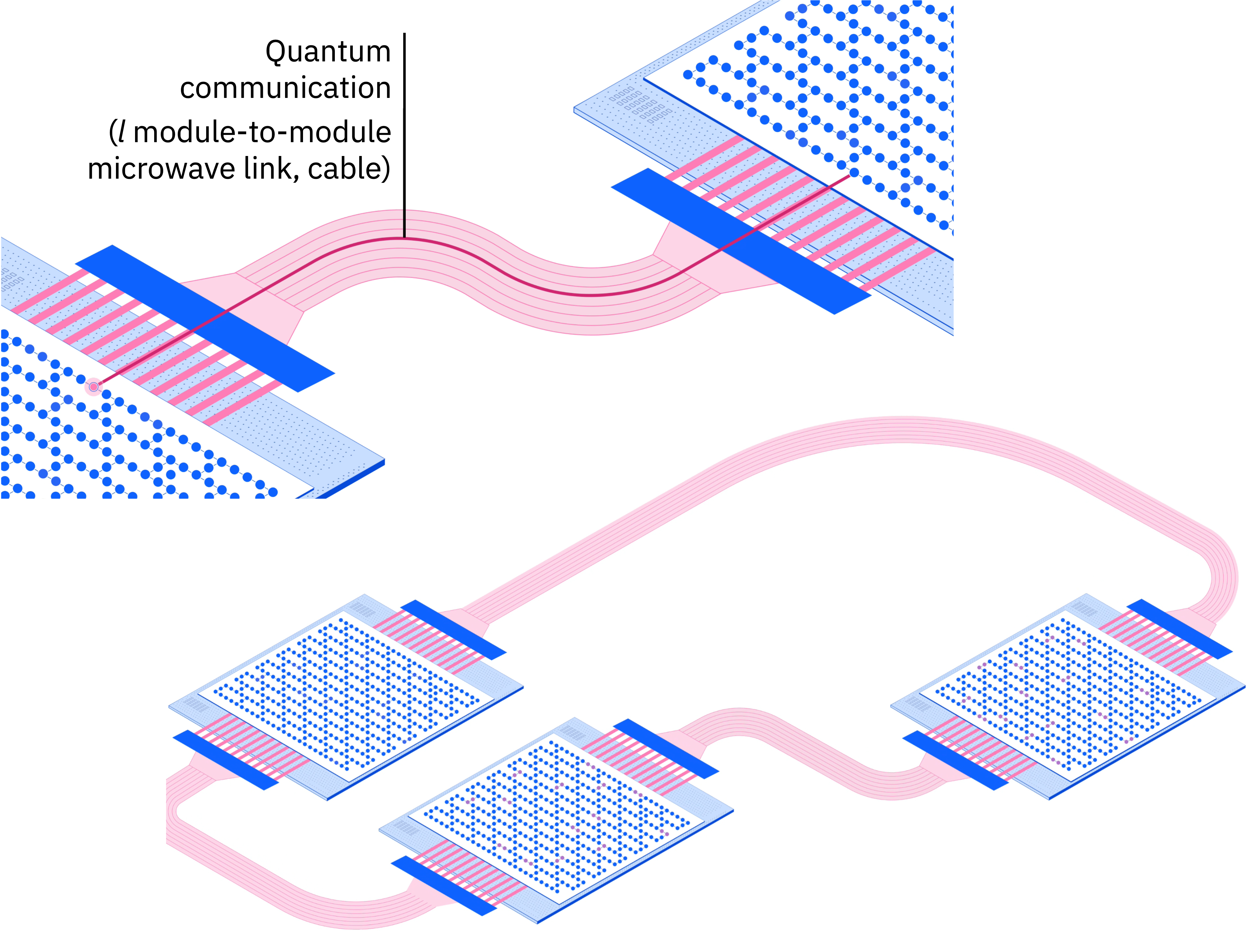}}
    \subfigure[{\it l}, {\it m}, {\it p} schemes can be combined to extend the scale of hardware to thousands of qubits.]{\includegraphics[width=0.49\textwidth]{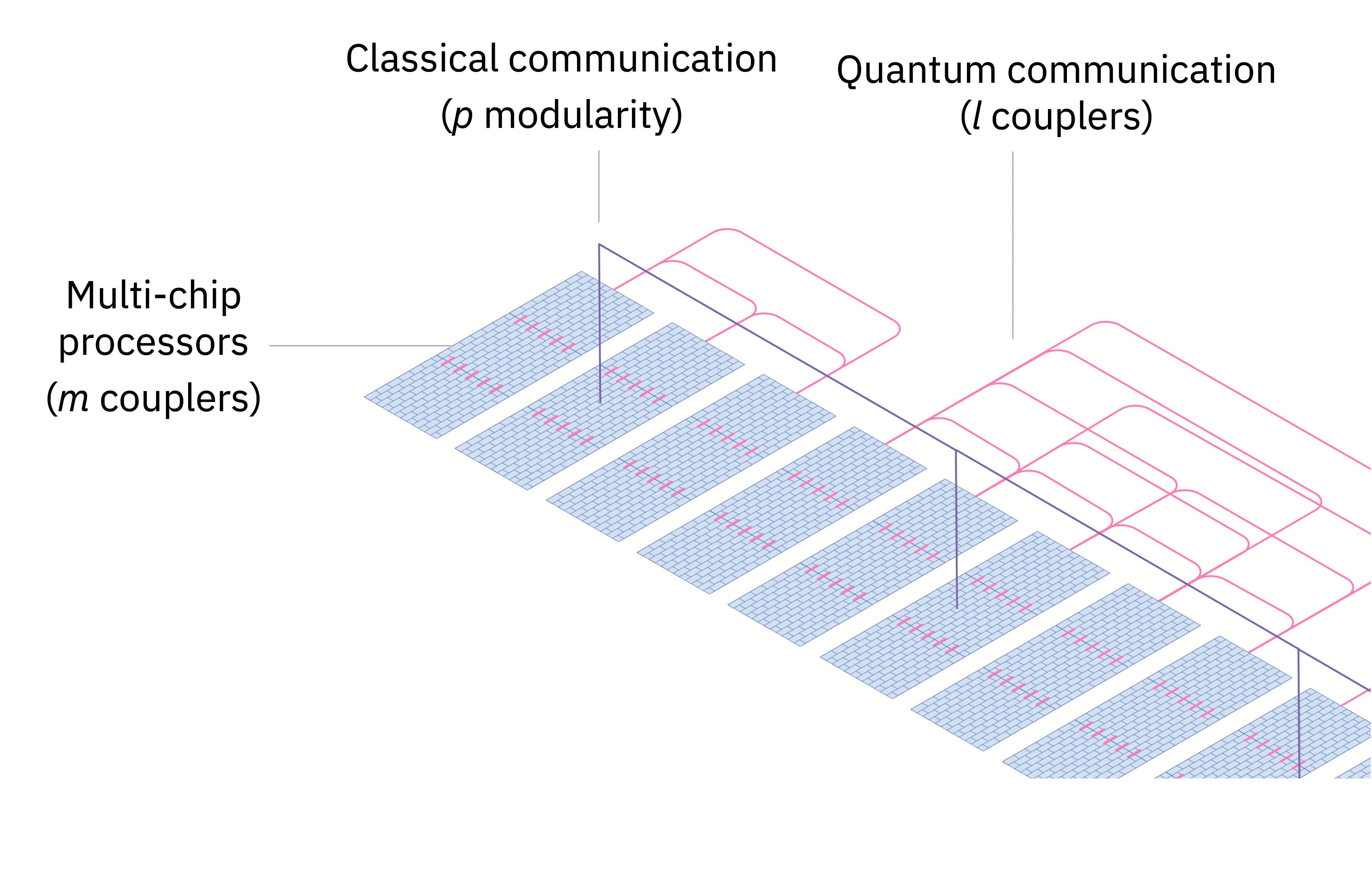}}
    \\ \vspace{0.3cm}
    \subfigure[{\it t} type modularity involves microwave-to-optical transduction to link QPUs in different dilution refrigerators.]{\includegraphics[width=0.5\textwidth]{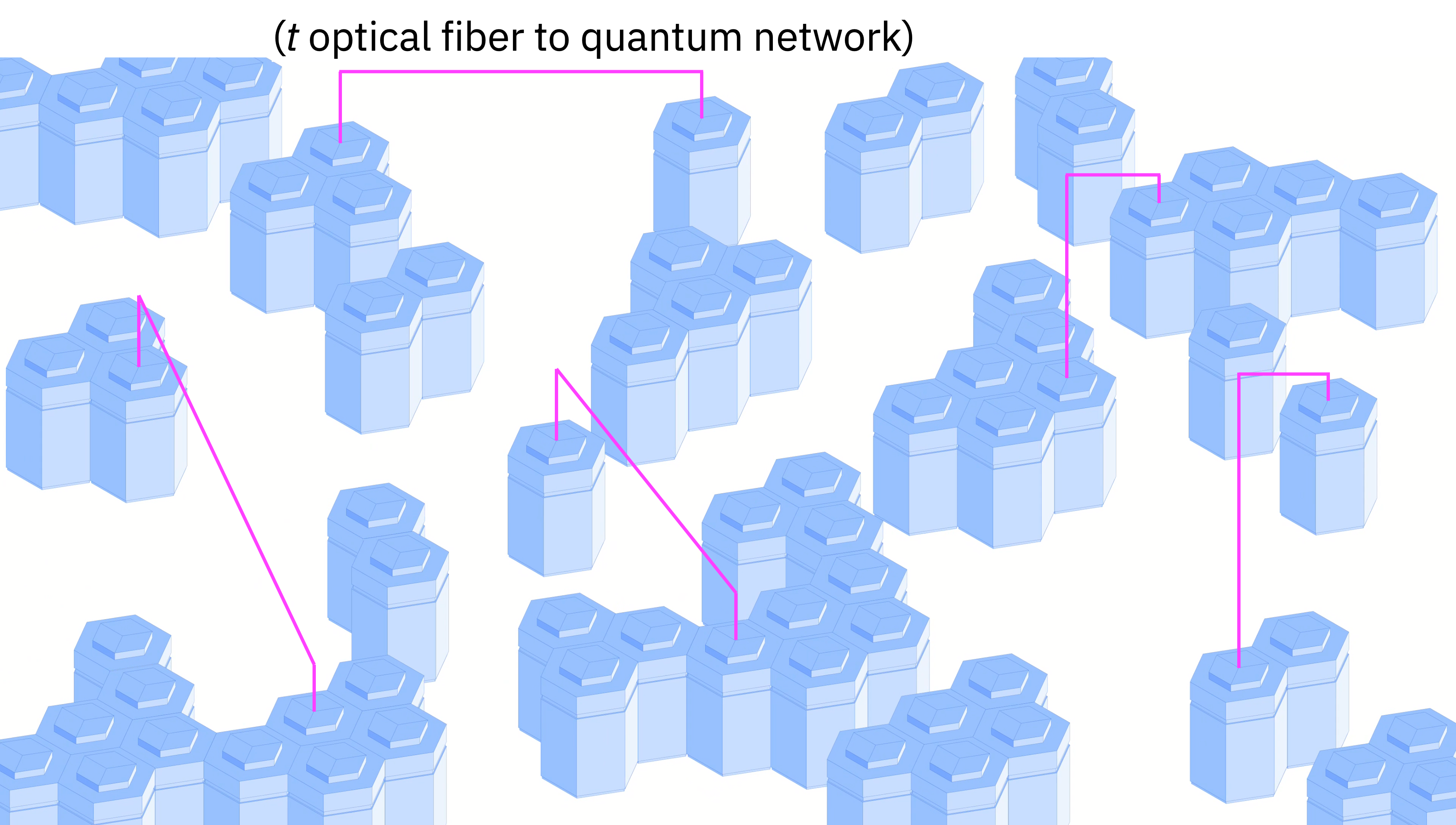}}
\caption{Beyond classical parallelization of QPUs, shown in (a), long-range quantum connections carry a high penalty in gate speed and fidelity.  As shown in (b)-(e), a high fidelity, large quantum system will likely involve three levels of modularity—a very short-range modularity {\it m} that allows breaking a QPU into multiple chips with minimal cost in gate speed and fidelity, a longer range connection {\it l} for use within a single cryogenic environment to both get around I/O bottlenecks and allow non-trivial topologies or routing, and a very long-range optical ``quantum network'' {\it t} to allow nearby QPUs to work together as a single quantum computational node (QCN). We will also need on-chip non-local couplers {\it c} as shown in (b) for the exploration of LDPC codes. In this figure, pink lines represent quantum communication and purple lines represent classical communication. } 
\label{mod}
\end{figure*}

The introduction of modular quantum systems will be key to bootstrapping ourselves from near-term quantum advantage towards long-term error-corrected quantum systems. These are systems with repeating unit cells that can be replaced if defective, with quantum links between the chips to entangle unit cells or perform remote gates.  This approach simplifies QPU design and test, and allows us to scale quantum systems at will.

\begin{table*}
    \centering
    \begin{tabular}{c @{\qquad} @{\quad} l @{\qquad} @{\quad} l}
    \toprule
    Type & Description & Use \\
    \hline
    $p$ & Real-time classical communication & Classical parallelization of QPUs \\
    $m$ & Short range, high speed, chip-to-chip & Extend effective size of QPUs \\
    $l$ & Meter-range, microwave, cryogenic & Escape I/O bottlenecks, enabling multi-QPUs \\
    $c$ & On-chip non-local couplers & Non-planar error-correcting code \\
    $t$ & Optical, room-temperature links & Ad-hoc quantum networking    \\
    \botrule
    \end{tabular}
    \caption{Types of modularity in a long-term scalable quantum system}
    \label{tab:modularity}
\end{table*}

In the near term, given limited or no error correction, the unit cells will require high-bandwidth and high fidelity links to connect them—there is not enough time to use complex protocols such as entanglement distillation. The simplest proposals to accomplish this extend quantum busses off chip, allowing the same gates between distant chips as on a single processor\cite{2102.13293, conner2021superconducting}. This ``dense modularity'', which we denote $m$, effectively extends the chip size.  This requires linking adjacent chips with ultra low loss, low cross-talk lines that are short enough to be effectively single-mode—the distance between chips has to be of the order of the distance between qubits on a single chip. Several technologies from classical computational hardware may be adaptable to this problem
but adding the flexibility to replace individual units will require other alternatives\cite{Larsson_2006}.

The high density of qubits in this ``dense modularity'' creates a spatial bottleneck for classical I/O and cooling. Proposals to ameliorate this near term include the development of high-density connectors and cables to route classical signals on and off the chip\cite{Tuckerman2016,Reilly2015}, and the addition of time- and frequency-domain multiplexing of controls.  A longer term approach to address this is to improve qubit connectivity through the use of a modified gate performed over a long conventional cable\cite{Zhong2021, Kurpiers2018,Leung2019}, called $l$ modularity.  Beyond allowing us to escape control and cooling bottlenecks, these long-range couplers enable the realization of non-2D topologies, thereby not only reducing the average distance between qubits but also opening the door to the exploration of more efficient non-2D LDPC error correction codes\cite{Campbell2017}. Developing these long-range couplers thus not only allows us to scale our near-term systems, but begins to form the basis for how to build quantum systems with mulitple QPUs.  

The technologies that enable both dense modularity and long-range couplers, once developed and optimized, will ultimately be ported back into the qubit chip to enable non-local, non-2D connectivity.  These on-chip non-local $c$ couplers will ultimately allow implementation of high-rate LDPC codes, bringing our long-term visions to completion.

Finally, connecting multiple quantum computers in an ad-hoc way will allow us to create larger systems as needed.  In this ``quantum networking" approach, the signals are typically envisioned to leave the dilution refrigerator, enabled by long-term technological advancements in microwave-to-optical transduction using photonic $t$ links between different fridges. 

With these four forms of modularity, we can redefine ``scale" for a quantum system by 
\[
\mathrm{n} = \left( [(q \, m)  \, l] t \right) p
\] 
where $n$ is the number of qubits in the entire modular and parallelized quantum system. The system is comprised of QPUs made from $m$ chips, each QPU having $q\times m$ qubits. The QPUs can be connected with $l\,t$ quantum channels (quantum parallelization), with $l$ of them being microwave connections and $t$ optical connections. Finally, to enable things like circuit cutting and speeding up error mitigation, each of these multi-chip QPUs can support classical communication, allowing $p$ classical parallelizations. 

A practical quantum computer will likely feature all five types of modularity—classical parallelization, dense chip-to-chip extension of 2D lattices of qubits ($m$), sparse connections with non-trivial topology within a dilution refrigerator ($l$), non-local on-chip couplings for error correction ($c$), and long-range fridge-to-fridge quantum networking ($t$) (\autoref{tab:modularity}). The optimal characteristic size of each level of modularity is an open question. The individual ``chip-to-chip" modules will still be made as large as possible, maximizing fidelity and connection bandwidth.  Performing calculations on a system like this with multiple tiers of connectivity is still a matter of research and development\cite{Nickerson2013, monroe2014modular}. 

Modularity needs to happen not just at the scale of the QPU, but at all levels of the system.  Modular classical control systems allow for easy subsystem testing, replacement, and assembly.  It's much easier to build a test infrastructure for a large number of small modules each year than a single, re-workable monolith.  The same can be said of refrigeration, with the added benefit that shipping and deploying monolithic large refrigeration systems is impractical.  A large number of our current failure points come in I/O and signal delivery, so modular solutions where sub-assemblies can be swapped out are essential.  The challenge here is moving the replaceable unit from a single unit (a cable) to a larger unit (a flexible ribbon cable or other cable assembly).

While the jury is still out on module size and other hardware details, what is certain is that the utility of any quantum computer is determined by its ability to solve useful problems with a quantum advantage while its adoption relies on the former plus our ability to separate its use from the intricacies of its hardware and physics-level operation. Ultimately, the power provided by the hardware is accessed through software that must enable flexible, easy, intuitive programming of the machines.

\begin{figure*}
\centering
    \subfigure{\includegraphics[width=\textwidth]{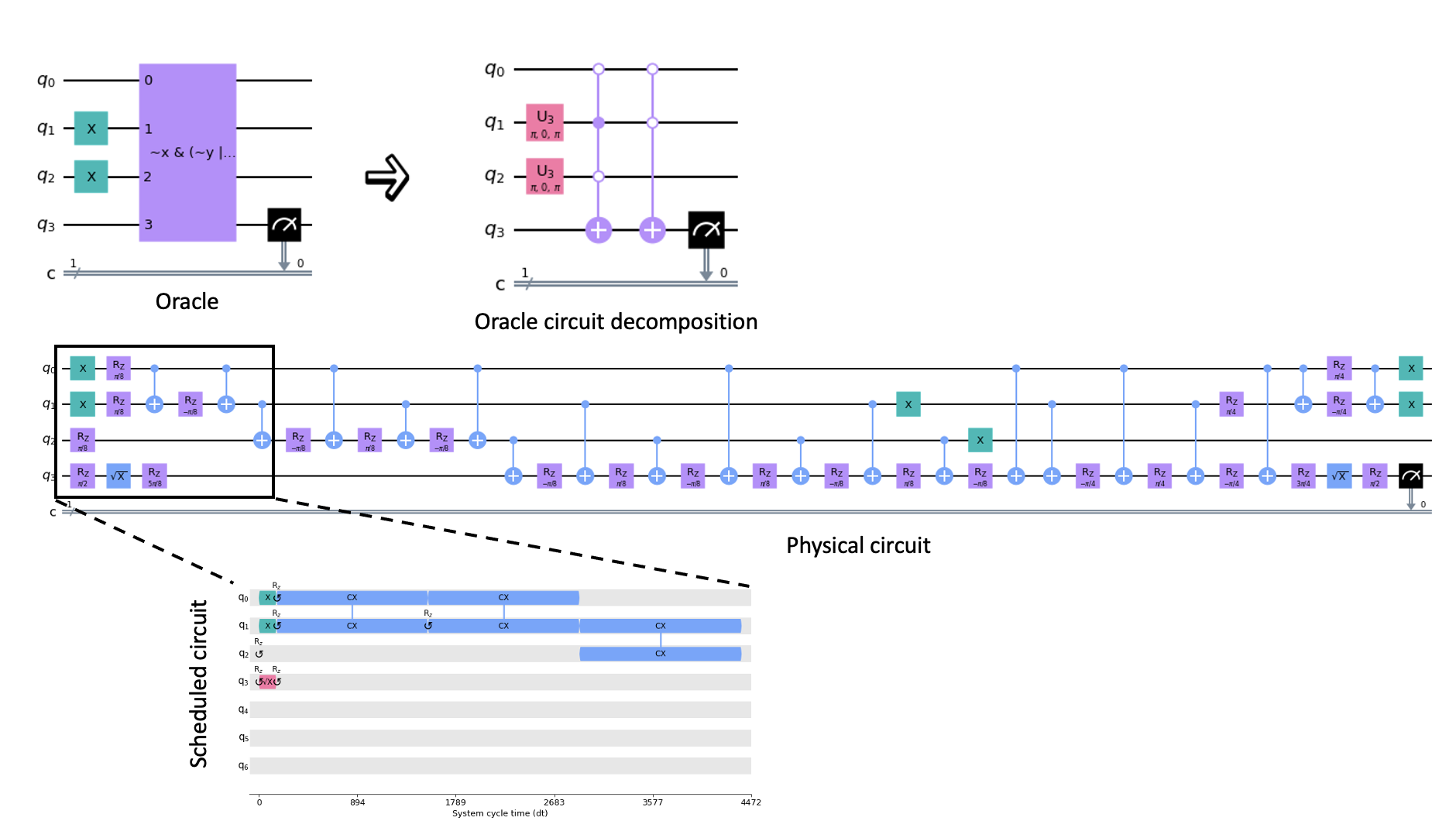}}
    \vspace{0.3cm}
    \hrule
    \vspace{0.3cm}
    \subfigure{\includegraphics[width=\textwidth]{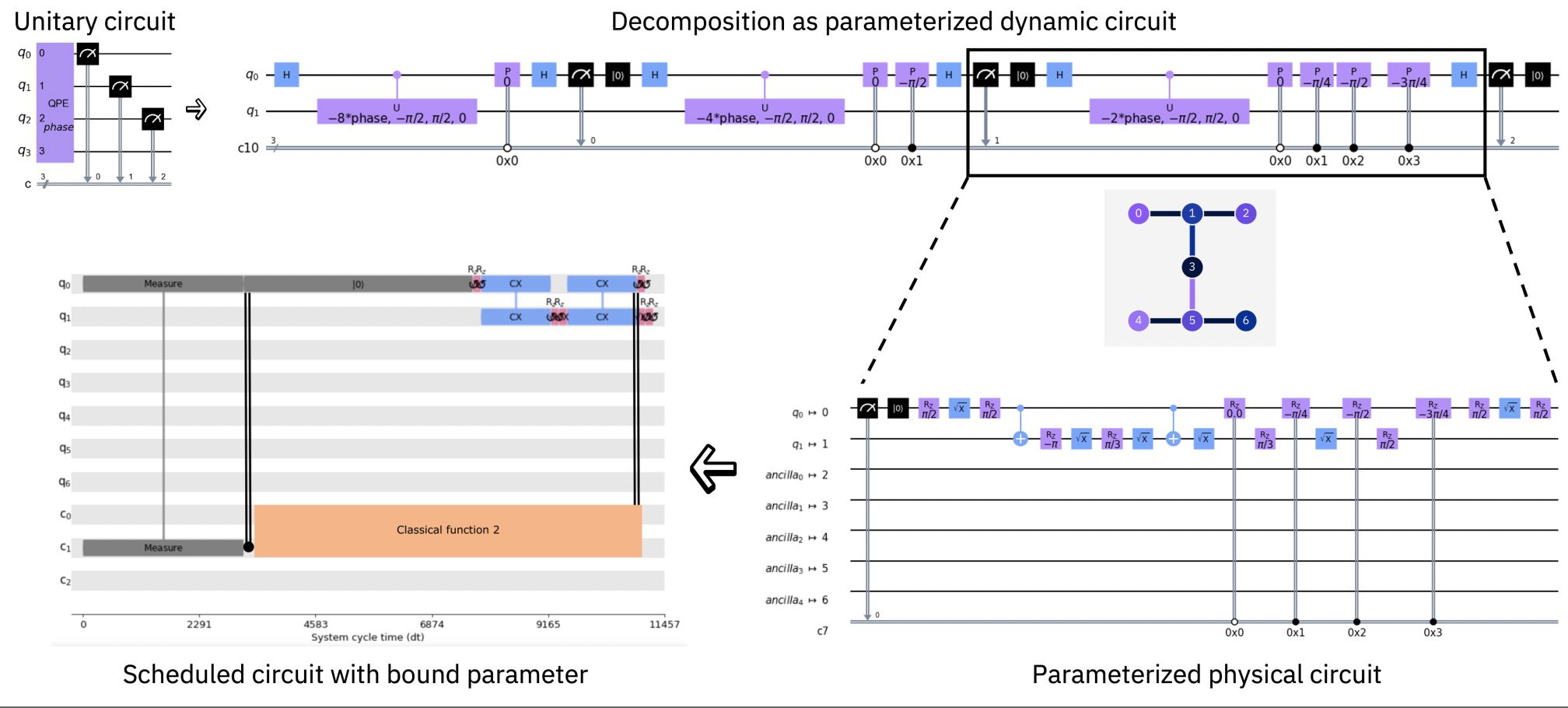}}
    \caption{Circuits can be represented at various levels. Unitary blocks represent circuits from libraries. These can be decomposed into parameterized circuits using the universal set of gates. Parameterized physical circuits use the physical gates supported by the hardware, while scheduled circuits specify timing, calibrations, and pulse shapes.} 
    \label{circuits}
\end{figure*}

\section{The Quantum Stack}
\label{sec:software}

For quantum computing to succeed in changing what it means to compute, we need to change the architecture of computing. Quantum computing is not going to replace classical computing but rather become an essential part of it. We see the future of computing being a quantum-centric supercomputer where QPUs, CPUs, and GPUs all work together to accelerate computations. In integrating classical and quantum computation, it is important to identify (1) latency, (2) parallelism (both quantum and classical), and (3) what instructions should be run on quantum vs. classical processors. These points define different layers of classical and quantum integration. 

Before we go into the stack, we need to redefine a quantum circuit. Here we define a quantum circuit as follows:  

\vspace{0.3cm}
\hrule
\vspace{0.15cm}
\noindent {\it A quantum circuit is a computational routine consisting of coherent quantum operations on quantum data, such as qubits, and concurrent (or real-time) classical computation. It is an ordered sequence of quantum gates, measurements, and resets, which may be conditioned on and use data from the real-time classical computation. If it contains conditioned operations, we refer to it is as a dynamic circuit.  It can be represented at different levels of detail, from defining abstract unitary operations down to setting the precise timing and scheduling of physical operations.} 
\vspace{0.15cm}
\hrule
\vspace{0.3cm}

This is general enough to represent the circuit model\cite{DiVincenzo95}, the measurement model\cite{Nielsen03}, and the adiabatic model\cite{Farhi2000} of computation, and special routines such as teleportation. Furthermore, it can represent the circuit at various levels: unitary (unitary block that could represent circuit libraries such as quantum phase estimation, classical functions, etc.), standard decomposition (reduced to a universal set of gates or expressing the classical functions as reversible gates), parameterized physical circuits (using the physical gates supported by the hardware, possibly including ancilla qubits not used in the circuit, or parameters that are easy to update in real-time), and scheduled circuits (complete timing information, calibrated gates, or gates with assigned pulse shape) (see fig. \ref{circuits}). OpenQASM\cite{qasm3} is an example intermediate representation for this extended quantum circuit and can represent each of these various abstractions. 

\begin{figure}[htb]
    \centerline{
    \includegraphics[width=0.45\textwidth]{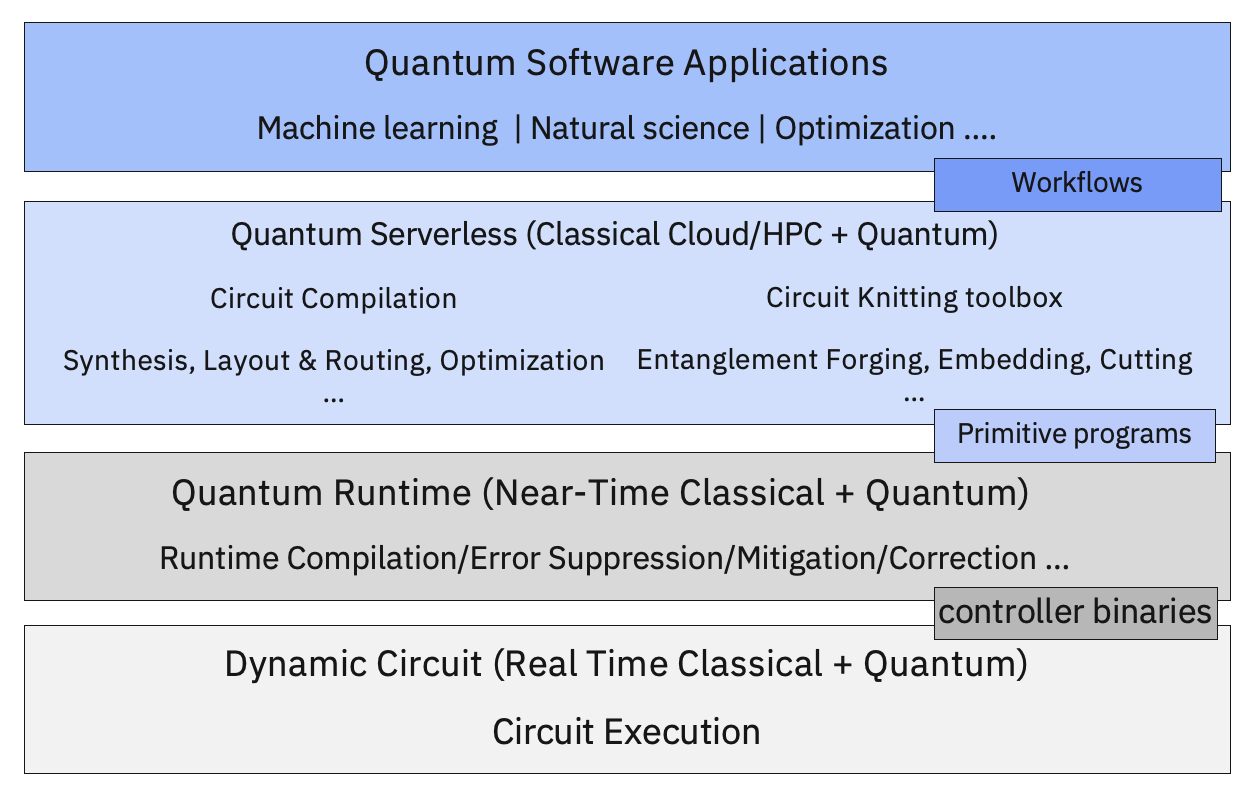}}
    \caption{The quantum software stack is comprised of four layers, each targeting the most efficient execution of jobs at different levels of detail. The bottom layer focuses on the execution of quantum circuits. Above it, the quantum runtime efficiently integrates classical and quantum computations, executes primitive programs, and implements error mitigation or correction. The next layer up (quantum serverless) provides the seamless programming environment that delivers integrated classical and quantum computations through the cloud without burdening developers with infrastructure management. Finally, the top layer allows users to define workflows and develop software applications.} 
\label{stack}
\end{figure}

With this extended quantum circuit definition, it is possible to define a software stack.  Fig. \ref{stack} shows a high level view of the stack, where we have defined four important layers: dynamic circuits, quantum runtime, quantum serverless, and software applications. At the lowest level, the software needs to focus on executing the circuit. At this level, the circuit is represented by controller binaries that will be very dependent on the superconducting qubit hardware, supported conditional operations and logic, and the control electronics used. It will require control hardware that can move data with low latency between different components while maintaining tight synchronization. For superconducting qubits, real-time classical communication will require a latency of $\sim$100 nanoseconds. To achieve this latency, the controllers will be located very close to the QPU. Today, the controllers are built using FPGAs to provide the flexibility needed, but as we proceed to larger numbers of qubits and more advanced conditional logic, we will need ASICs or even cold CMOS. 

\begin{figure*}
    \centerline{
    \includegraphics[width=0.85\textwidth]{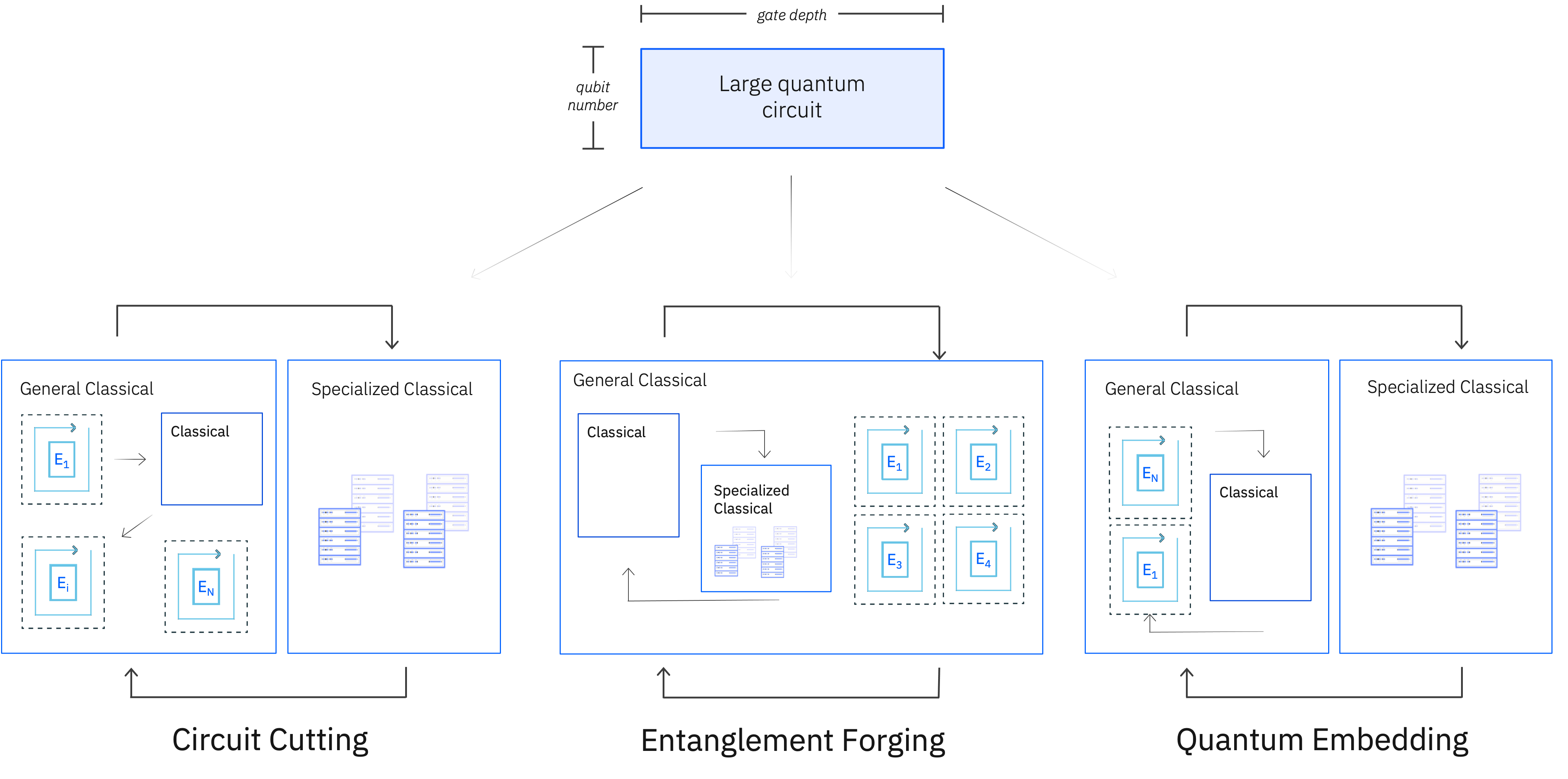}}
    \caption{Example of a quantum serverless architecture integrating quantum and classical computations.  Quantum runtimes are illustrated by estimator primitives. Cloud computing is illustrated by general classical computing. Specialized classical computing such as high precision computing (HPC) or graphics processing units (GPUs) could be integrated into the serverless architecture. In circuit cutting, a larger circuit is split into many smaller circuits using a specialized classical computer. For each of the smaller circuits, an estimator primitive is executed ($E_1$, $\cdots$, $E_N$) and if needed, a classical computing routine could be used to condition future circuits on the results of previous estimators. The process can be repeated as needed. In entanglement forging, a 2N-qubit wavefunction is decomposed into a larger number of N-qubit circuits. The entanglement synthesis may need to be offloaded to specialized classical processors. For each N-qubit circuit, an estimator $E_N$ is executed and combined to give the global outcome. This process could be repeated if used in a variational algorithm. Quantum embedding separates sub-parts of a problem that can be simulated classically from those computationally most costly and requiring quantum computations. A specialized classical computer could be used to condition the problem on previous outcomes. The quantum simulations employ estimators $E_N$ running on QPUs. The estimators can condition quantum circuits on previous outcomes with classical calculations run on the general classical processors. Collectively, this set of tools allows larger systems to be simulated with higher accuracy.} 
\label{knitting}
\end{figure*}

We refer to the next level up as the quantum runtime layer. This is the core quantum computing layer. In the most general form, we expect a quantum computer to run quantum circuits and generate non-classical probability distributions at their outputs. Consequently, much of the workloads are sampling from or estimating properties of distributions. The quantum runtime thus needs to include at least two primitive programs: the sampler and the estimator. The sampler collects samples from a quantum circuit to reconstruct a quasi-probability distribution of the output. The estimator allows users to efficiently calculate expectation values of observables. 

The circuit sent to the runtime would be a parameterized physical circuit. The software would perform a runtime compilation and process the results before returning the corrected outcome. The runtime compilation would update the parameters, add error suppression techniques such as dynamical decoupling, perform time-scheduling and gate/operation parallelization, and generate the controller code. It would also process the results with error mitigation techniques, and in the future, error correction. Both today's error mitigation and tomorrow's error correction will place strong demands on the classical computing needed inside these primitive programs. The circuit execution time could be as low as 100 microseconds (maybe even 1 microsecond for error correction), which is not possible over the cloud. It will need to be installed as part of the quantum computer. Fortunately, error mitigation is pleasingly parallel, thus using multiple QPUs to run a primitive will allow the execution to be split and done in parallel.

At the third level, we imagine software that can combine advanced classical calculations with quantum calculations.  As described earlier in this paper, introducing classical computing can enable ideas such as circuit kniting. Here we need to be able to call quantum primitive programs as well as perform classical calculations such as circuit partitions. We call this a workflow (fig. \ref{knitting} shows examples of workflows for circuit knitting). We refer to quantum serverless as the software architecture and tooling that supports this in a way that allows developers to focus only on code and not on the classical infrastructure. Along with circuit knitting, this layer will also allow advanced circuit compiling that could include synthesis, layout and routing, and optimization—all of which are parts of the circuit reduction that should happen before sending the circuit to execute.

Finally, at the highest level of abstraction, the computing platform must allow users to efficiently develop software applications. These applications may need access to data and to resources not needed by the quantum computation itself but needed to provide the user an answer to a more general problem. 

Each layer of the software stack we just described brings different classical computing requirements to quantum computing and defines a different set of needs for different developers. Quantum computing needs to enable at least three different types of developers: kernel, algorithm, and model developers. Each developer creates the software, tools, and libraries that feed the layers above, thereby increasing the reach of quantum computing. 

The kernel developer focuses on making quantum circuits run with high quality and speed on quantum hardware. This includes integrating error suppression, error mitigation, and eventually, error correction into a runtime environment that returns a simplified application programming interface (API) to the next layer. 

\begin{figure*}[htb]
    \centerline{
    \includegraphics[width=0.75\textwidth]{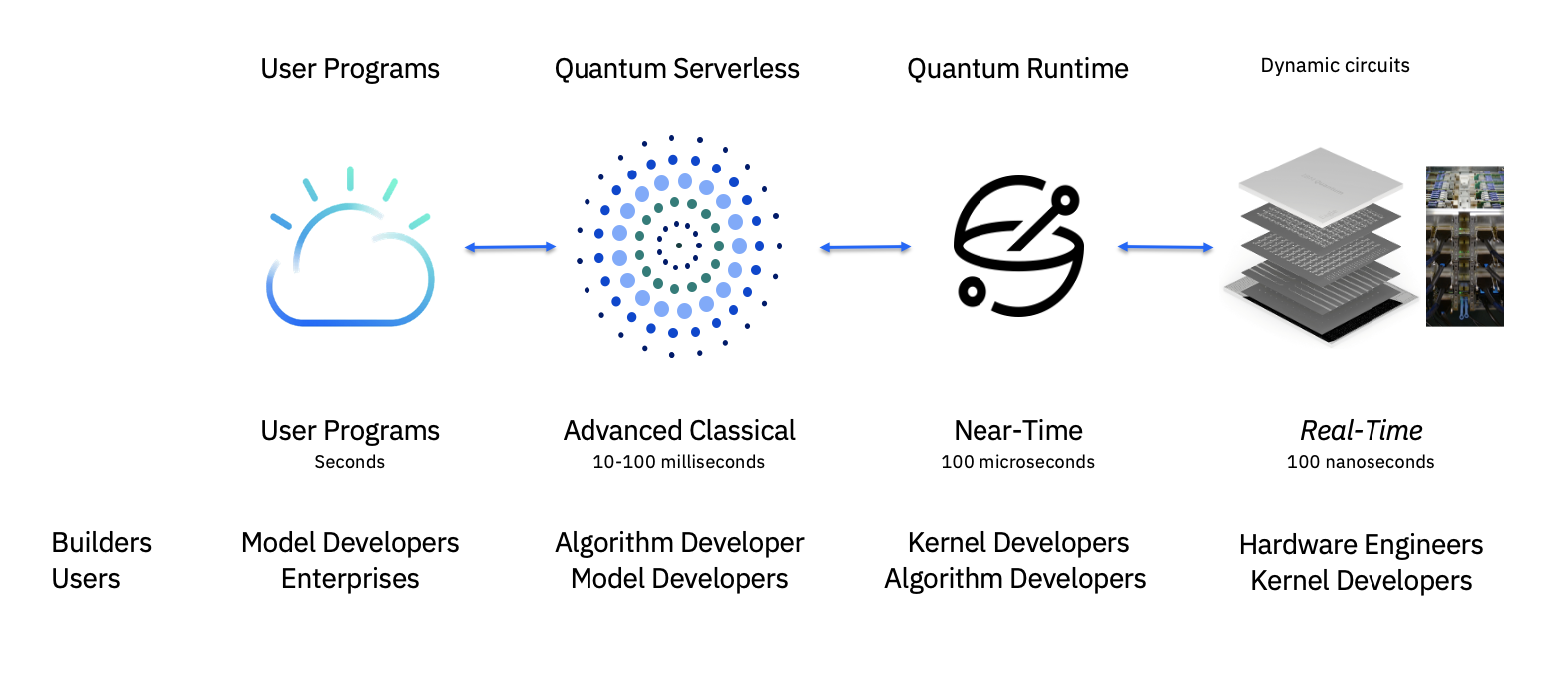}}
    \caption{The time scales and resources involved in quantum computing depend on the needs of the different types of developers and the level of abstraction at which they work. Quantum researchers and kernel developers work closer to the hardware while model developers require the highest level of software abstraction.} 
\label{timescales}
\end{figure*}

\begin{figure}
    \centerline{
    \includegraphics[width=0.5\textwidth]{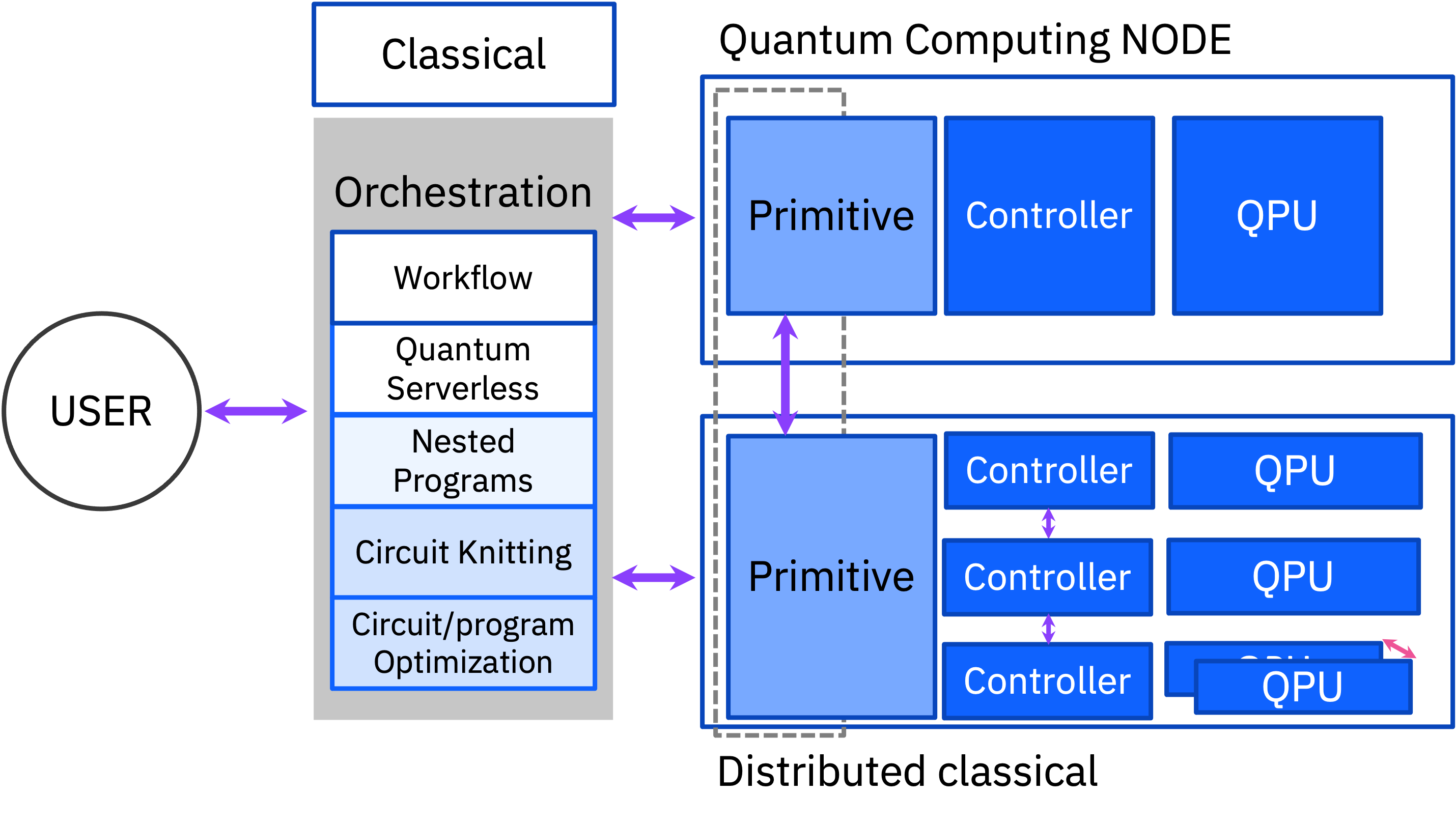}}
    \caption{Model of a cluster-like architecture integrating classical processors with QPUs to address latency, parallelization, and the distribution of instructions among classical and quantum processors. The darker the color, the lower the latency required.} 
\label{cluster}
\end{figure}

The algorithm developer combines quantum runtime with classical computing, implements circuit knitting, and builds heuristic quantum algorithms and circuit libraries. The purpose is to enable quantum advantage. Finally, as we demonstrate examples of quantum advantage, the model developer will be able to build software applications to find useful solutions to complex problems in their specific domain, enabling enterprises to get value from quantum computing. Fig. \ref{timescales} summarizes the types of developers addressed by each layer of the software stack and the time scales involved depending on the type of job being executed and how close to the hardware each developer is working.

In putting all of this together and scaling to what we call a quantum-centric supercomputer, we do not see quantum computing integrating with classical computing as a monolithic architecture. Instead, fig. \ref{cluster} illustrates an architecture for this integration as a cluster of quantum computational nodes coupled to classical computing orchestration. The darker the color, the closer the classical and quantum nodes must be located to reduce latency.  Threaded runtimes can execute primitives on multiple controllers. Classical communication in real time between the controllers can be used  to enable things like circuit cutting. The figure also shows how future QPUs with quantum parallelization ($l$ and $t$ couplers) can be controlled by a single controller. We imagine that there could be workloads that need near-time classical communication (i.e., calculations based on the outcome of circuits that must complete in around 100 microseconds) or to share states between the primitives, enabled by a data fabric. Finally, the orchestration would be responsible for workflows, serverless, nested programs (libraries of common classical+quantum routines), the circuit knitting toolbox, and circuit compilation.

\section{Conclusion}

In conclusion, we have charted how we believe that quantum advantage in some scientifically relevant problems can be achieved in the next few years. This milestone will be reached through (1) focusing on problems that admit a super-polynomial quantum speedup and advancing theory to design algorithms—possibly heuristic—based on intermediate depth circuits that can outperform state-of-the-art classical methods, (2) the use of a suite of error mitigation techniques and improvements in hardware-aware software to maximize the quality of the hardware results and extract useful data from the output of noisy quantum circuits, (3) improvements in hardware to increase the fidelity of QPUs to 99.99\% or higher, and (4) modular architecture designs that allow parallelization (with classical communication) of circuit execution. Error mitigation techniques with mathematical performance guarantees, like PEC, albeit carrying an exponential classical processing cost, provide a mean to quantify both the expected run time and the quality of processors needed for quantum advantage. This is the near-term future of quantum computing.

Progress in the quality and speed of quantum systems will improve the exponential cost of classical processing required for error mitigation schemes, and a combination of error mitigation and error correction will drive a gradual transition toward fault-tolerance. Classical and quantum computations will be tightly integrated, orchestrated, and managed through a serverless environment that allows developers to focus only on code and not infrastructure. This is the mid-term future of quantum computing.

Finally, we have seen how realizing large-scale quantum algorithms with polynomial run times to enable the full range of practical applications requires quantum error correction, and how error correction approaches like the surface code fall short of the long term needs owing to their inefficiency in implementing non-Clifford gates and poor encoding rate. We outlined a way forward provided by the development of more efficient LDPC codes with a high error threshold, and the need for modular hardware with non-2D topologies to allow the investigation of these codes. This more efficient error correction is the long-term future of quantum computing.

\begin{acknowledgments}
We thank Kristan Temme, Abhinav Kandala, Ewout van den Berg, Jerry Chow, Antonio C\'{o}rcoles, Ismael Faro, Blake Johnson, Tushar Mittal, and Matthias Steffen for their assistance reviewing this manuscript.
\end{acknowledgments}

\bibliography{qc_perspectives_IBMQuantum}

\end{document}